\documentclass[reprint,twocolumn,superscriptaddress, amssymb,aps,prl]{revtex4-2}

\usepackage{graphicx}
\usepackage{xcolor} 
\usepackage{amsmath}

\usepackage{physics}

\begin{document}
\title{Probing Coherent Many-Body Spin Dynamics in a Molecular Tweezer Array Quantum Simulator} 
\author{Yukai Lu}
\affiliation{Department of Physics, Princeton University, Princeton, New Jersey 08544 USA}
\affiliation{Department of Electrical and Computer Engineering, Princeton University, Princeton, New Jersey 08544 USA}
\author{Connor M. Holland}
\affiliation{Department of Physics, Princeton University, Princeton, New Jersey 08544 USA}
\author{Callum L. Welsh}
\affiliation{Department of Physics, Princeton University, Princeton, New Jersey 08544 USA}
\author{Xing-Yan Chen}
\affiliation{Department of Physics, Princeton University, Princeton, New Jersey 08544 USA}
\author{Lawrence W. Cheuk}
\email{lcheuk@princeton.edu}
\affiliation{Department of Physics, Princeton University, Princeton, New Jersey 08544 USA}

\begin{abstract}
Models of interacting quantum spins are used in many areas of physics ranging from the study of magnetism and strongly correlated materials to quantum sensing. In this work, we study coherent many-body dynamics of interacting spin models realized using polar molecules trapped in rearrangeable optical tweezer arrays. Specifically, we encode quantum spins in long-lived rotational states and use the electric dipolar interaction between molecules, together with Floquet Hamiltonian engineering, to realize $1/r^3$ XXZ and XYZ models. We microscopically probe several types of coherent dynamics in these models, including quantum walks of single spin excitations, the emergence of magnon bound states, and coherent creation and annihilation of magnon pairs. Our results establish molecular tweezer arrays as a new quantum simulation platform for interacting quantum spin models.
\end{abstract}
\maketitle

\section{Introduction}
Interacting quantum spin models play an important role in physics. Not only are they used to describe magnetism~\cite{Auerbach1994magnetismbook}, but they also have connections to a broad range of topics ranging from quantum sensing~\cite{Pezze2018RMPMetrology} to gravitational physics~\cite{Sachdev1993sy,kitaev2015syk}. While theoretical investigations of these models date back nearly a century~\cite{Bethe1931}, recent experimental advances have enabled quantum simulations of these models with a variety of platforms. These include ultracold atoms in optical lattices ~\cite{Bloch2008ManyBodyReview}, neutral atom arrays~\cite{Browaeys2020Review}, trapped ions~\cite{Monroe2021Review}, superconducting circuits~\cite{Salathe2015SuperconductingSpin}, atomic cavity QED systems~\cite{Periwal2021CavitySpin}, and polar molecules~\cite{Bohn2017Review}. 

In the past two decades, polar molecules have emerged as a platform particularly well-suited for studying dipolar interacting systems~\cite{Carr2009review,Bohn2017Review,Blackmore2018QuantumReview}. The rich internal structure of molecules provides long-lived internal states to encode quantum spins while electric dipolar interactions between molecules allow long-ranged ($1/r^3$) and anisotropic interactions. Already, the long-lived rotational states have been used to encode spins~\cite{Yan2013Spin, gregory2021robust, burchesky2021rotcoh, Christakis2023Spin, Holland2023SE, Bao2023SE, Picard2025SE, Ruttley2025SE}, and experiments with ultracold molecular gases trapped in optical lattices have realized various interacting spin models, including itinerant ones~\cite{Yan2013Spin, Christakis2023Spin, Miller2024XYZ,Carroll2025SpintJ}. Nevertheless, these experiments have lacked microscopic control, and until recently~\cite{Christakis2023Spin}, also microscopic detection at the single molecule level. Such capabilities are highly desirable for the study of many-body physics. They allow one to assemble many-body systems from the ground up, initialize out-of-equilibrium many-body states, and detect spatial correlations that often sensitively probe emergent phenomena such as magnetic and hidden order.

Recently, polar molecules trapped in rearrangeable optical tweezer arrays have emerged as a platform that combines the richness of molecules with microscopic control and detection. Since the first demonstrations of trapping single molecules in tweezers~\cite{Anderegg2019Tweezer, Holland2023Bichro, Zhang2021NaCsArray, Ruttley2023tweezer}, coherent dipolar interactions between  pairs of molecules have been observed~\cite{Holland2023SE, Bao2023SE,Picard2025SE,Ruttley2025SE}, providing the key building block for interacting quantum spin models. Nevertheless, due to challenges in preparing molecular tweezer arrays with sufficient fidelity and size, before our work, the interacting many-body regime had been out of reach, with previous experiments limited to two interacting molecules~\cite{Holland2023SE, Bao2023SE, Picard2025SE, Ruttley2025SE}.

In this work, we probe coherent dynamics of interacting spin models in the many-body regime with molecular tweezer arrays for the first time. Specifically, we leverage recent advances in internal state preparation~\cite{Holland2025Erasure} to create mesoscopic one-dimensional (1D) tweezer arrays of polar CaF molecules with sufficiently low defect rates in both occupation and internal state. We encode a spin-1/2 degree of freedom in two long-lived rotational states and use the native intermolecular electric dipolar interactions, together with Floquet Hamiltonian engineering~\cite{Eckardt2017FloquetReview, Choi2020Floquet,Weidemuller2021Floquet, Scholl2022Floquet,Christakis2023Spin, Florian2023Magnon, Miller2024XYZ, Lei2025Floquet}, to realize tunable $1/r^3$ XXZ and XYZ spin models. We then leverage our microscopic control and detection capabilities to explore several highly out-of-equilibrium phenomena in these models. These include quantum walks of single spin excitations, dynamics of magnon bound states, and coherent creation and annihilation of magnon pairs.

\begin{figure}[h!]
	{\includegraphics[width=\columnwidth]{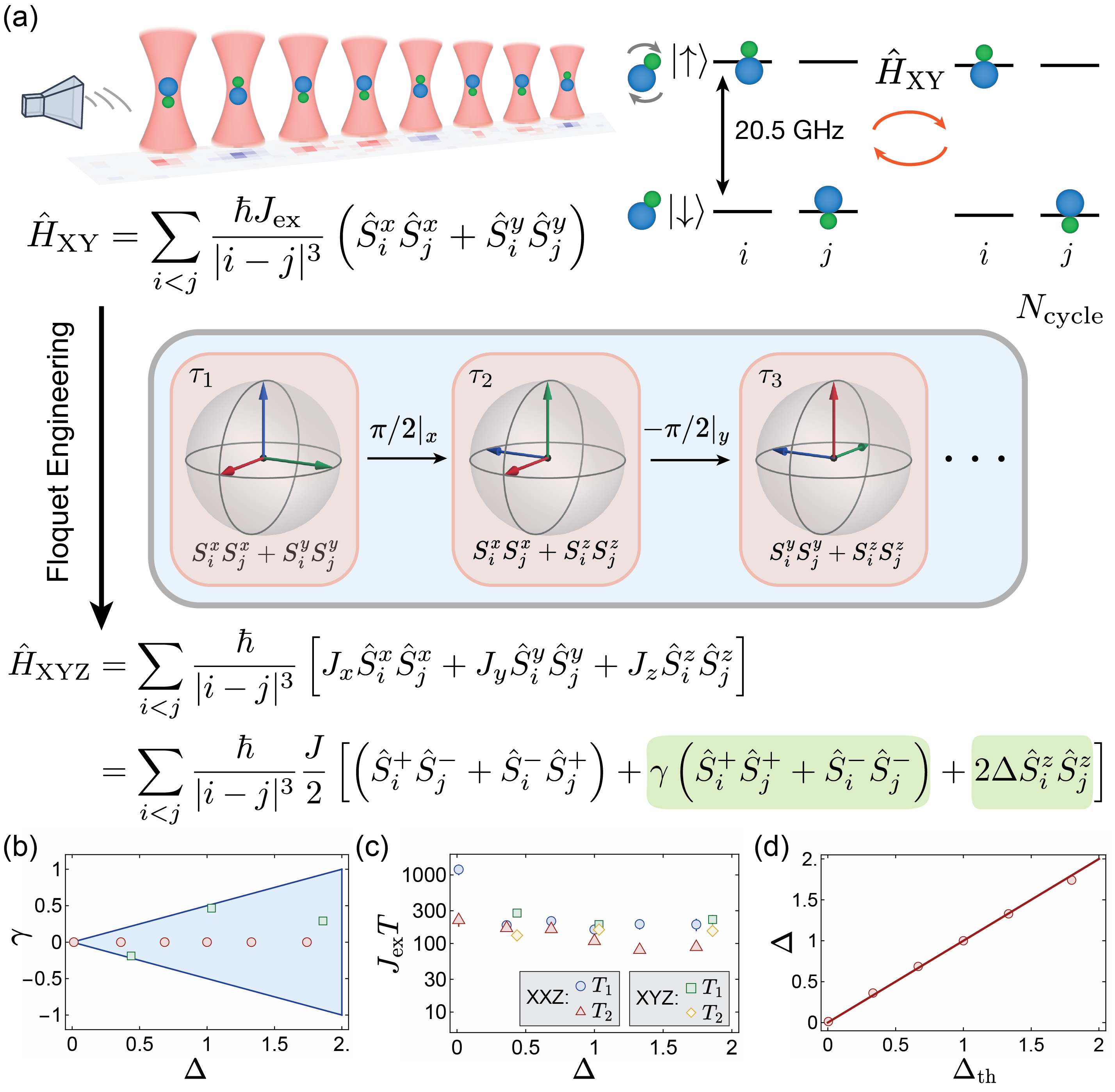}}
	\caption{\label{fig_1} Realizing $1/r^3$ XXZ/XYZ Spin Models in a Molecular Tweezer Array. (a) Polar CaF molecules are laser-cooled and trapped in a rearrangeable optical tweezer array. A spin-1/2 degree of freedom is encoded in two molecular rotational states and electric dipolar interactions provide a native $1/r^3$ XY Hamiltonian $\hat{H}_{\rm XY}$. Floquet engineering using  microwave pulses converts $\hat{H}_{\rm XY}$ into $\hat{H}_{\rm XYZ}$. Specifically, $\pi/2$ pulses toggle the instantaneous spin frame, providing two additional terms: (1) an Ising term $\hat{S}_i^z \hat{S}_j^z$ with strength $\Delta$, and (2) a pair creation/annihilation interaction term $\hat{S}_i^+ \hat{S}_j^++\hat{S}_i^- \hat{S}_j^-$ with strength $\gamma$. (b) The blue shaded region indicates the accessible parameter space of $\gamma$ and $\Delta$. The red (green) data points show the parameters explored in this work. (c) Normalized depolarization time $J_{\rm ex}T_1$ (blue circles) and decoherence time $J_{\rm ex}T_2$ (red triangles) for the XXZ sequences, together with $J_{\rm ex}T_1$ (green squares) and $J_{\rm ex}T_2$ (yellow diamonds) for the XYZ sequences. The data is corrected for blackbody leakage. These times are generally on the scale of $10^2$. (d) Experimentally measured Ising strength $\Delta$ with isolated molecular pairs versus the theoretically predicted value $\Delta_{\text{th}}$. The red line indicates $\Delta = \Delta_{\text{th}}$.
	}
	\vspace{-0.2in}
\end{figure}

\section{Realizing Tunable $1/r^3$ XXZ/XYZ Models via Floquet Hamiltonian Engineering}
Our work begins with laser-cooled CaF molecules individually trapped in a 1D tweezer array~\cite{Holland2023Bichro,Holland2023SE,Holland2025Erasure}. Quantum spins are encoded in two long-lived molecular states $\ket{\downarrow}=\ket{X, v=0, N=0, J=1/2, F=1, m_F=0}$ and $\ket{\uparrow}=\ket{X, v=0, N=1, J=1/2, F=0, m_F=0}$, where $X$, $v$, and $N$ are the electronic, vibrational, and rotational quantum numbers, respectively. These states, from neighboring rotational manifolds, interact via the electric dipolar interaction, natively providing spin-exchange interactions described by the $1/r^3$ XY spin Hamiltonian:
\begin{eqnarray}
\hat{H}_{\rm{XY}} &=& \sum_{i<j}\frac{\hbar J_{\text{ex}}}{|i-j|^3}(\hat{S}_i^x\hat{S}_j^x+\hat{S}_i^y\hat{S}_j^y)\nonumber \\
&=& \sum_{i<j}\frac{\hbar J_{\text{ex}}/2}{|i-j|^3}(\hat{S}_i^+\hat{S}_j^- + \hat{S}_i^-\hat{S}_j^+),
\end{eqnarray}
where $\hat{S}_i^\beta$ are spin-1/2 operators for molecule $i$. For our work, the molecules are separated by 2.10(3)\,$\mu{\rm m}$,  with antiferromagnetic interactions of strength $J_{\text{ex}} \approx 2\pi  \times 32.93(8)\,\text{Hz}$.

To extend the class of accessible Hamiltonians, we make use of Floquet Hamiltonian engineering~\cite{Choi2020Floquet}. In brief, we apply a periodic sequence of microwave pulses on the $\ket{\uparrow}$-$\ket{\downarrow}$ transition interspersed with free evolution times (Fig.~\ref{fig_1} (a)). This rapidly toggles the instantaneous spin axes, allowing new interaction terms of the form $\hat{S}^\beta_i \hat{S}^\beta_j$ whose strengths are tunable through pulse sequence parameters. When the sequence period $T$ is short compared to the interaction timescale ($|J_{\rm{ex}} T|\ll 1$), the system effectively evolves under the $1/r^3$ XYZ Hamiltonian:
\begin{eqnarray}
\hat{H}_{\text{XYZ}} &=& \sum_{i<j} \frac{\hbar}{|i-j|^3} \left[ J_x \hat{S}_i^x\hat{S}_j^x +J_y \hat{S}_i^y\hat{S}_j^y +J_z \hat{S}_i^z \hat{S}_j^{z} \right] \nonumber \\
&=& \sum_{i<j} \frac{\hbar J}{|i-j|^3} \left[ \frac{1}{2}(\hat{S}_i^+\hat{S}_j^- + \hat{S}_i^-\hat{S}_j^+) \right. \nonumber \\
& &\left.+ \frac{\gamma}{2}(\hat{S}_i^+\hat{S}_j^+ + \hat{S}_i^-\hat{S}_j^{-})+\Delta \hat{S}_i^z \hat{S}_j^z\right]. 
\label{eq:XYZ} 
\end{eqnarray}
Compared to $\hat{H}_{\rm{XY}}$, $\hat{H}_{\rm{XYZ}}$ contains two new terms: (1) an Ising interaction $\hat{S}^z_{i}\hat{S}_{j}^z$ with strength $\Delta$, and (2) a pair creation/annihilation interaction $\hat{S}^+_{i}\hat{S}_{j}^+ + \hat{S}^-_{i}\hat{S}_{j}^-$ with strength $\gamma$ (Fig. \ref{fig_1}(a)). Since Floquet engineering redistributes interactions among the terms in Eq.~(\ref{eq:XYZ}), the accessible parameter range is constrained by $J_x+J_y+J_z=2J_\text{ex}$ with $0\leq J_\alpha\leq J_\text{ex}$ ($\alpha=x,y,z$). Equivalently, the accessible region is described by $J(1+\Delta/2) =J_{\text{ex}}$ with $0\leq\Delta\leq2$ and $|\gamma|\leq \Delta/2$ (Fig.~\ref{fig_1}(b)). We note that our platform is particularly well-suited for Floquet engineering. Microwave pulses ($\mu\mathrm{s}$-scale) can be applied much faster than the interaction timescale $1/J_{\text{ex}}\approx 30\,\text{ms}$, allowing us to use short Floquet periods $T$ that satisfy $|J_\text{ex}T|\ll 1$. For our work, $J_{\rm{ex}} T\sim 0.15$ and we use robust pulse sequences where higher-order corrections only enter at $(J_{\rm{ex}} T)^2 \sim 0.02$.

A key challenge in Floquet Hamiltonian engineering is ensuring that pulse errors are small enough so that the target Hamiltonian is faithfully realized over the timescale of interest. Deeply satisfying the Floquet condition, in fact, implies that a large number of pulses ($\sim 1/(J_\text{ex}T)\gg 1$) is needed to reach the interaction timescale of $1/J_\text{ex}$. To verify that our Floquet engineering performs sufficiently well in our system, we first measure single particle depolarization and decoherence errors for each pulse sequence used. We initialize molecules at large separations of $16.8\,\mu\mathrm{m}$ with negligible interactions, and measure their $1/e$ spin depolarization time $T_1$ and decoherence time $T_2$ for empirically optimized XXZ and XYZ sequences~\cite{Supplement}. For all parameters explored, we obtain $J_{\rm ex}T_1, J_{\rm ex} T_2 \sim \mathcal{O}(10^2)$ (Fig.~\ref{fig_1}(c)), substantially longer than the interaction timescale of $1/J_\text{ex}$.

We next verify that the target Hamiltonian is faithfully realized and calibrate its parameters using isolated pairs of interacting molecules. We initialize the molecules into different product states and observe their subsequent spin dynamics under Floquet driving, which allows us to directly calibrate $J$, $\gamma$, and $\Delta$~\cite{Supplement}. As an example, for XXZ Hamiltonians ($\gamma=0$), Fig.~\ref{fig_1}(d) shows that the measured $\Delta$ agrees well with the expected value $\Delta_\text{th}$. Similar measurements are performed for XYZ sequences ($\gamma\neq0$) (Fig.~\ref{fig_1}(b)), where we also find good agreement with theory, indicating that we have faithfully engineered $\hat{H}_\text{XYZ}$. 

\section{$1/r^3$ XXZ Model: Observing Quantum Walks in the Single Magnon Sector}
Having characterized our Floquet engineering performance with one and two particles, we next investigate $\hat{H}_\text{XYZ}$ in the \textit{many-body} setting. We prepare mesoscopic 1D chains of $N_{\text{mol}}=7$ and $N_{\text{mol}}=8$ molecules, the largest systems practically accessible in our experiment. We first focus on the XXZ case ($\gamma=0$), where an underlying $U(1)$ symmetry leads to the conservation of the total spin along $\hat{z}$, $\hat{S}^z=\sum_i \hat{S}_i^z$. The Hilbert space is partitioned into disconnected sectors with fixed $S^z$.

We investigate dynamics in the smallest non-trivial sector ($|S^z |=N_{\text{mol}}/2-1$), the single magnon sector, which contains states with a single spin excitation on top of a polarized background. Experimentally, we prepare $N_{\text{mol}}=7$ molecules in $\ket{\downarrow\downarrow\downarrow\uparrow\downarrow\downarrow\downarrow}$, which has a single spin up excitation. We then measure the site-resolved $\ket{\downarrow}$ population $P^{\downarrow}_i(t)$ after an evolution time $t$. To suppress depolarization errors, we post-select experimental shots where six $\ket{\downarrow}$ molecules are detected~\cite{Google2022PhotonBoundState}. Correspondingly, $P^{\uparrow}_i = 1- P^{\downarrow}_i$ then reveals the location of the $\ket{\uparrow}$ excitation.

\begin{figure*}
	{\includegraphics[width=2\columnwidth]{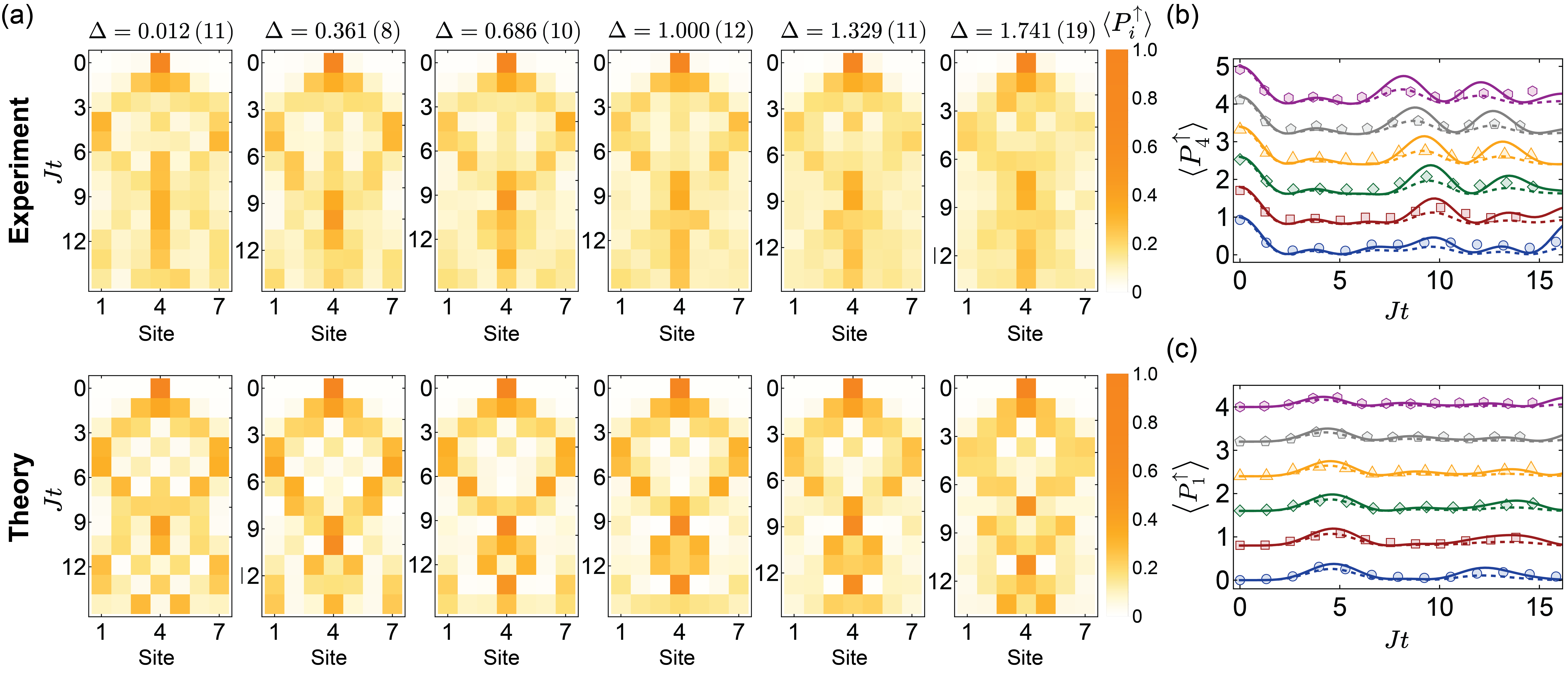}}
	\caption{\label{fig_2} 
		Single Magnon Dynamics in the $1/r^3$ XXZ Model. (a) Evolution of $\langle P_i^\uparrow\rangle$ at Ising strengths $\Delta=0.012(11), 0.361(8), 0.686(10), 1.00(12), 1.329(11), 1.741(19)$ versus time $J t$, where $J$ is the nearest-neighbor spin-exchange strength. The initially localized $\ket{\uparrow}$ spin excitation performs a coherent quantum walk. Top row shows experimental data, which are qualitatively consistent with exact diagonalization simulations shown in the bottom row. (b) Spin up probability of the center site $\langle P_4^\uparrow\rangle$ versus time $Jt$ (c) Spin up probability of the leftmost site $\langle P_1^\uparrow\rangle$ versus time $Jt$. For (b,c), solid lines show numerical simulations of the ideal model. Dashed lines include a phenomenological exponential damping envelope with damping constant obtained by fitting to $\langle P_4^\uparrow\rangle$. The data and corresponding fits are ordered by increasing $\Delta$ from bottom to top and vertically offset by 0.8 for clarity. The values of $\Delta$ are the same as (a).}
	\vspace{-0.2in}
\end{figure*}

Fig.~\ref{fig_2}(a) shows $P^{\uparrow}_i(t)$ over the range of accessible Ising strengths $\Delta$. We observe that the $\ket{\uparrow}$ excitation undergoes a coherent quantum walk driven by the spin-exchange terms $\hat{S}^+_i\hat{S}^-_j +\hat{ S}^-_i \hat{S}^+_j$. Our observations agree qualitatively with exact-diagonalization simulations of the ideal $1/r^3$ XXZ model, albeit with reduced contrast. We attribute this reduction to interaction disorder in $J_{\rm{ex}}$ from the thermal motion of the molecules, rather than decoherence or higher-order Floquet terms~\cite{Supplement}. Quantitatively, Fig.~\ref{fig_2}(b,c) show the observed population in the middle site and leftmost site for various $\Delta$. We find quantitative agreement with exact diagonalization simulations once a phenomenological exponential damping envelope is included to capture interaction disorder. These observations establish that coherent dynamics persist up to a timescale of $\sim 10/J$, surviving reflections from the boundaries of the chain.

\section{$1/r^3$ XXZ Model: Observing Magnon Bound States and Their Dynamics}
\begin{figure*}[h!]
	{\includegraphics[width=1.5\columnwidth]{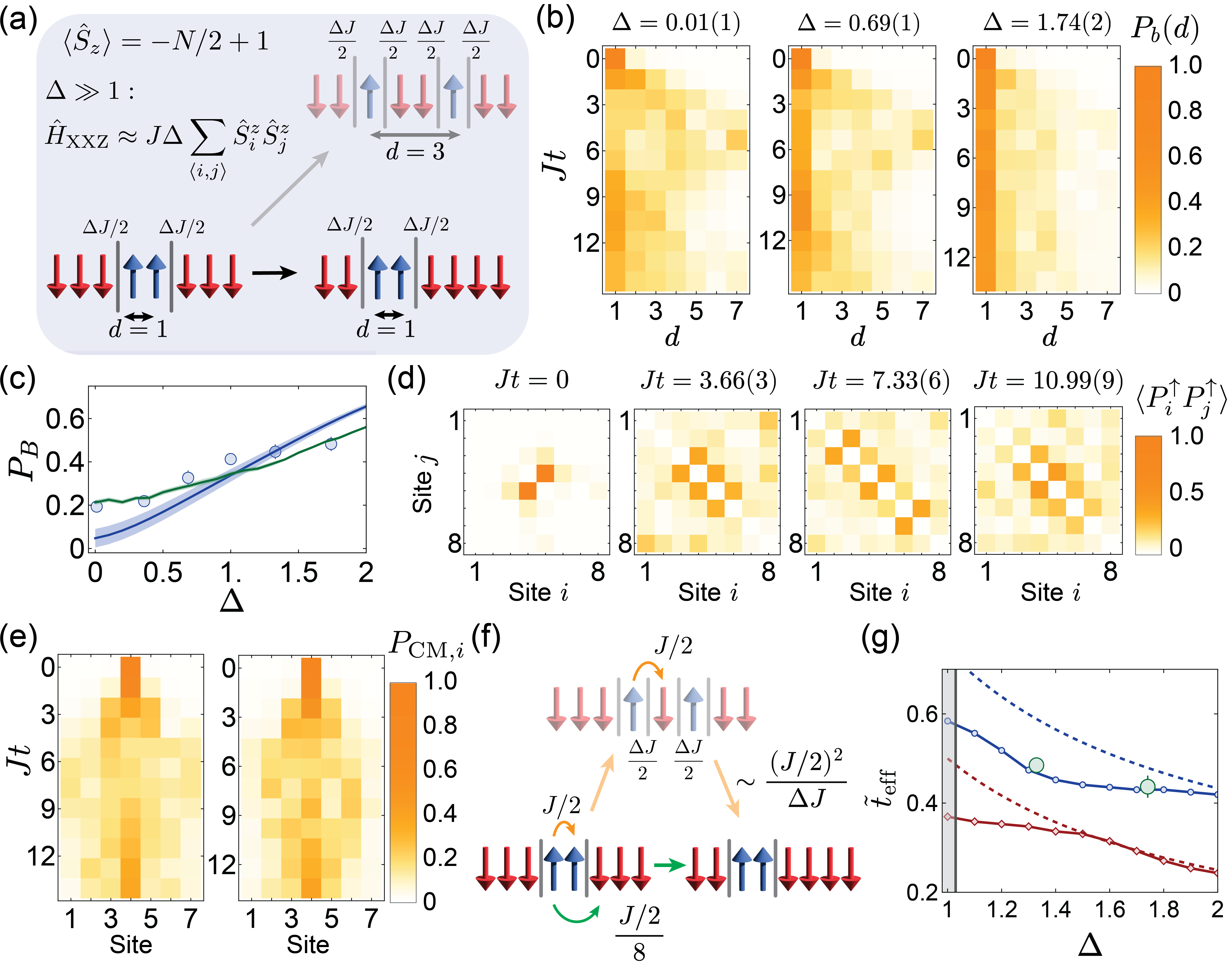}}
	\caption{\label{fig_3} Two-Magnon Dynamics in the $1/r^3$ XXZ Model. (a) For nearest-neighbor interactions and $\Delta\gg1$, breaking a spin pair creates domain walls with energy $\Delta J$. Energy conservation implies domain wall conservation, leading to magnon bound states. (b) $\ket{\uparrow}$-$\ket{\uparrow}$ separation probability $P_b(d,t)$ versus time $t$. (c) $P_{B}$, the equilibrium value of $P_b(1,t)$, versus $\Delta$. The green (blue) curve shows exact diagonalization results with (without) interaction disorder and state preparation/detection infidelities, with the shaded region showing error bands from fitting~\cite{Supplement}. (d) Measured spatial correlators $\langle P^{\uparrow}_i P^{\uparrow}_j\rangle$ for $\Delta=1.741\,(19)$ versus time. Significant weight on the first off-diagonals indicates the two spins propagating as a bound pair. Color scales are normalized to the measured peak value. (e) Center-of-mass distribution $P_{{\rm CM}, i}(t)$ versus time for $ \Delta=1.329\,(11)$ (left) and $ \Delta=1.741\,(19)$ (right). Revival at the central site indicates coherent dynamics. (f) For $1/r^3$ interactions, pair hopping includes both a second-order and a first-order contribution. (g) Pair hopping rate ${\tilde{t}_{\rm{eff}}}$ versus $\Delta$. Green points show measured values for $\Delta$ above the thermodynamic bound state threshold ($\Delta\approx 1.03$)~\cite{Supplement}. The dashed red (blue) line shows perturbation theory for nearest-neighbor ($1/r^3$) interactions; the solid red (blue) line shows the exact-diagonalization result for nearest-neighbor ($1/r^3$) interactions. }
	\vspace{-0.2in}
\end{figure*}

We next explore the two-magnon sector ($| S^z |=N_{\text{mol}}/2-2$), which contains states with two spin excitations. Here, the interplay between Ising and spin-exchange interactions can lead to the formation of bound states of the excitations, i.e. magnon bound states. Qualitatively, this can be understood in the nearest-neighbor limit using a domain wall picture (Fig.~\ref{fig_3}(a)). In the strong Ising limit $\Delta \gg 1$, the eigenstates in a fixed $ S^z $ sector are approximate product states of $\ket{\uparrow}$ and $\ket{\downarrow}$, with energies determined by the domain wall number $N_D$. Energy conservation thus implies domain wall conservation. For an initial state with two adjacent spin excitations $\left|\uparrow\uparrow\right\rangle$, which has two domain walls, the two excitations must remain bound to conserve $N_D$. 

The appearance of magnon bound states in finite-range $1/r^\alpha$ XXZ models depends on $\alpha$, and has been experimentally studied across several platforms. They have been investigated in nearest-neighbor XXZ models ($\alpha \approx \infty$) with neutral atoms~\cite{Fukuhara2013Magnon,Weckesser2025EBH}, in related nearest-neighbor Floquet-XXZ models with superconducting qubits~\cite{Google2022PhotonBoundState}, and in long-range XXZ models ($\alpha\approx1$) with trapped ions~\cite{Florian2023Magnon}. Here, we experimentally investigate these bound states for $\alpha=3$, where the bound state band becomes fully isolated from the continuum in the thermodynamic limit when \(\Delta \gtrsim 1.03\)~\cite{Supplement}.

Experimentally, we initialize 8 molecules in $\ket{\downarrow\downarrow\downarrow\uparrow\uparrow\downarrow\downarrow\downarrow}$ and observe their subsequent evolution. For each experimental run, we measure $P^{\downarrow}_i(t)$, from which we extract $P_b(d,t)$, the probability that the two $\ket{\uparrow}$ spins are separated by $d$ sites. As shown in Fig.~\ref{fig_3} (b), increasing $\Delta$ leads to $P_b(d,t)$ being progressively localized near $d=1$, indicating that the $\ket{\uparrow}$ spins remain close to each other. Quantitatively, we extract the equilibrium value $P_B$ of $P_b(d=1,t)$ for evolution times up to $t\sim 6/J$, which acts as a proxy for bound state formation. We find that $P_B$ increases with Ising strength $\Delta$ (Fig.~\ref{fig_3}(c)), consistent with stronger binding and bound state formation. To provide further evidence of magnon bound states, we also examine the two-point correlators $\langle P^{\uparrow}_i P^{\uparrow}_j\rangle$~\cite{Fukuhara2013Magnon}. At large Ising strength $\Delta=1.741(19)$, significant spin correlations appear at $|i-j|=1$, indicating that spin excitations propagate together (Fig.~\ref{fig_3}(d)).

We next extract the propagation speed of the bound pairs by post-selecting for snapshots where the two spins in $\ket{\uparrow}$ are adjacent and constructing the distribution of their mean position $P_{{\rm CM},i}(t)$ (Fig.~\ref{fig_3}(e)). For the strongest Ising strengths ($\Delta = 1.329(11)$ and $\Delta=1.741(19)$), we observe reflection of the bound pair from the boundaries and its subsequent revival at the central site. This provides a clear signature of coherent quantum walk dynamics of the bound pair. 

By comparing with Fig.~\ref{fig_2}(a), we observe that the magnon bound pair propagates at a slower speed than the single free magnon. This speed is in fact a sensitive probe of next-nearest-neighbor interactions present in our $1/r^3$ models~\cite{Browaeys2025tJV}. To determine the magnon pair speed, we fit $P_{{\rm CM},i}(t)$ to an effective nearest-neighbor tight-binding model with hopping amplitude $t_{\rm{eff}}$~\cite{Supplement}. In the nearest-neighbor XXZ model at strong Ising strength $\Delta\gg 1$, pair motion arises from a second-order process: the front spin virtually hops forward, temporarily breaking the pair at an energy cost of $\sim\Delta J$, followed by a hop of the rear spin. The normalized hopping amplitude $\tilde{t}_{\rm eff}=2t_{\rm eff}/J$ is then $\sim 1/(2\Delta)$. However, with $1/r^3$ interactions, the rear excitation can directly hop over the front excitation via the next-nearest-neighbor coupling, giving rise to a first-order contribution of $1/8$ (Fig.~\ref{fig_3}(f))~\cite{Supplement}. Therefore, when $\Delta\gg1$, $\tilde{t}_{\rm{eff}}$ allows measurement of next-nearest-neighbor interactions~\cite{Browaeys2025tJV}. Fig.~\ref{fig_3}(g) shows the experimentally extracted hopping rate $\tilde{t}_{\rm{eff}}$ versus Ising strength $\Delta$. We find that the data agrees with exact simulations using an ideal $1/r^3$ XXZ model with 8 sites but disagrees with a nearest-neighbor XXZ model. Despite strong finite-size effects, the data unambiguously reveals next-nearest-neighbor interactions.

\section{$1/r^3$ XYZ Model: Coherent Creation and Annihilation of Magnon Pairs}
We next investigate coherent dynamics in the fully anisotropic XYZ model where $\gamma \neq 0$. While nearest-neighbor XYZ models have been extensively studied theoretically~\cite{Lieb1961XY, Baxter1972XYZ2}, experimental realizations have remained challenging. Only recently have XYZ spin models been realized with atoms and molecules~\cite{Weidemuller2021Floquet, Gross2023XYZ, Luo2025XYZ, Miller2024XYZ}. These experiments have so far accessed either few-spin systems (e.g., three spins) or mean-field dynamics. Here, we go beyond these regimes by microscopically probing highly out-of-equilibrium magnon pair creation and annihilation dynamics in chains of $N_{\text{mol}}=8$ spins.

\begin{figure*}[h!]
	{\includegraphics[width=2\columnwidth]{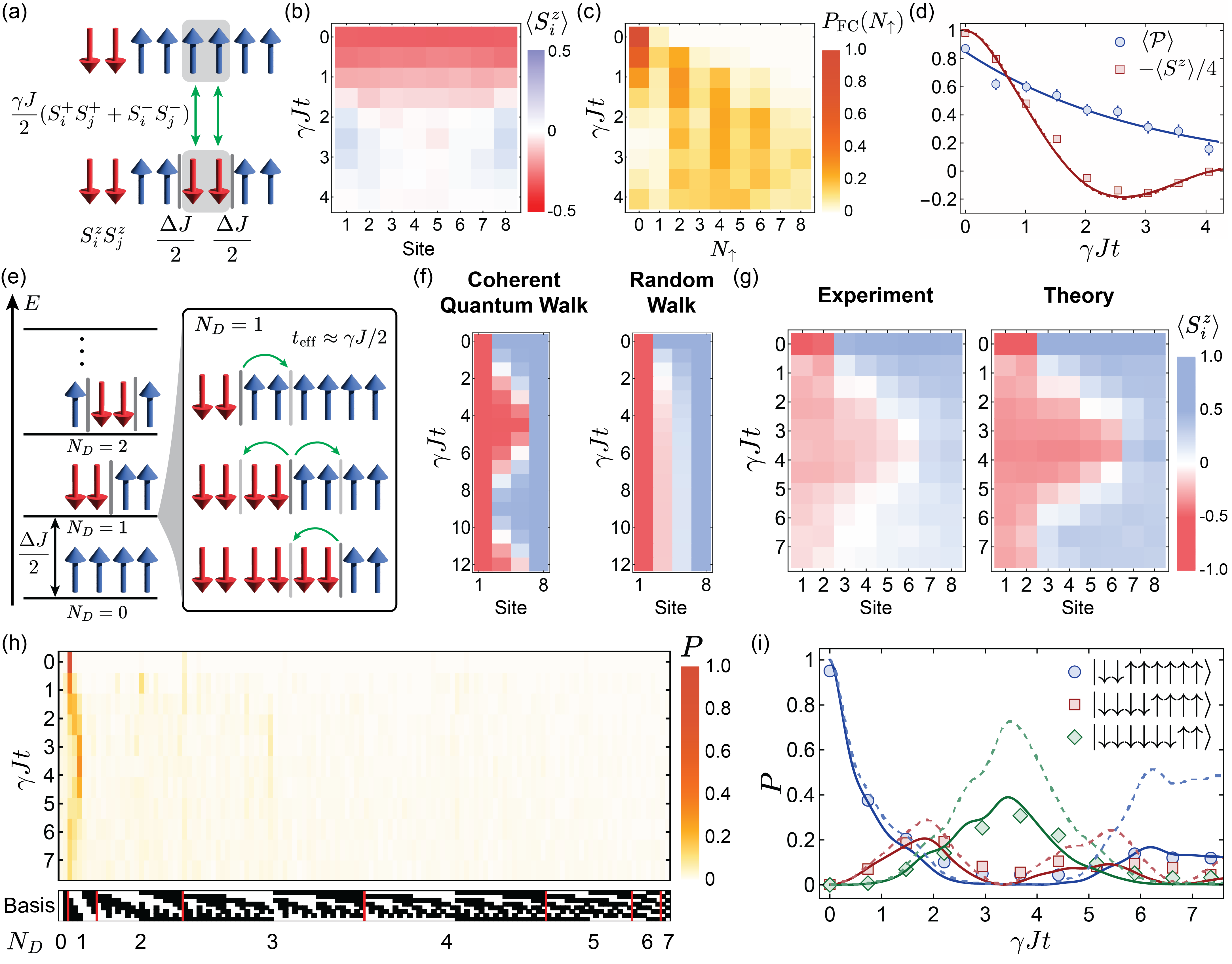}}
	\caption{\label{fig_4} Coherent Magnon Pair Creation and Annihilation in the $1/r^3$ XYZ Model. (a) For $\hat{H}_{\text{XYZ}}$ under a nearest-neighbor approximation, adjacent spins are flipped pairwise, creating domain walls with energy $J\Delta/2$. (b) $\langle \hat{S}_i^z\rangle$ versus time $\gamma Jt$ for an initial state $\ket{\downarrow}^{\otimes N_{\rm mol}}$, with $\Delta=1.033(11)$ and $\gamma=0.466(5)$. (c) Full counting statistics $P_\text{FC}(N_\uparrow)$ versus $\gamma Jt$. (d) The parity $\langle\mathcal{P}\rangle$ (blue circles) and total spin $-2 \langle \hat{S}^z\rangle/N_{\rm mol}$ (red squares) versus time. The blue solid line is an exponential fit to $\langle\mathcal{P}\rangle$ with a time constant of $9.6(8)/J_{\text{ex}}$. The red solid (dashed) line shows exact diagonalization simulations with (without) a fitted exponential envelope. (e) Energy diagram for $\hat{H}_{\text{XYZ}}$ with $\Delta\gg 1$ and nearest-neighbor approximation. Pair creation is mapped onto a quantum walk of a domain wall. (f) Simulated $\langle \hat{S}_i^z\rangle$ versus time for a coherent quantum walk (left) and an incoherent random walk (right). (g) $\langle \hat{S}_i^z\rangle$ versus time for the initial state $\ket{\downarrow\downarrow \uparrow \uparrow\uparrow\uparrow\uparrow\uparrow}$ with $\Delta=1.860\,(13)$ and $\gamma=0.290(3)$. (h) Measured microstate population $P$ versus time for data in (g). (i) Population $P$ of the 3 most occupied microstates versus time. Dashed lines are exact diagonalization results. Solid lines include an exponential envelope with a time constant of $36.8(4)/J_{\text{ex}}$.}
	\vspace{-0.2in}
\end{figure*}

A key difference of XYZ models, compared to XXZ models, is a broken $U(1)$ symmetry that leads to non-conservation of $\hat{S}^z$. Specifically, a new pair creation and annihilation term ${\hat{S}_i^+ \hat{S}_j^+ + \hat{S}_i^- \hat{S}_j^-}$ flips spins pairwise at a rate $\approx \gamma J/2$ (Fig.~\ref{fig_4}(a)). Consequently, the system only conserves the parity of the number of $\ket{\uparrow}$ molecules $N_{\uparrow}$, with the corresponding parity operator given by $\hat{\mathcal{P}}=(-1)^{\hat{N}_{\uparrow}}=(-1)^{N_{\text{mol}}}\prod_i (2\hat{S}_i^z)$.

To reveal parity conservation, we initialize molecules in the spin-polarized state $\ket{\downarrow}^{\otimes N_{\text{mol}}}$ and measure $\langle \hat{S}^z_i \rangle$ versus time $t$ for $\Delta=1.033(11)$ and $\gamma=0.466(5)$. To reject errors arising from vacant sites, which severely affect the parity, we perform full readout of both spin populations in every shot~\cite{Supplement}. This allows shots with vacancies to be post-selected out. Our data (Fig.~\ref{fig_4}(b)), post-selected on zero vacancies, shows that $\langle \hat{S}_i^z \rangle$ rapidly depolarizes on a timescale $t \approx 1/(J\gamma)$, much faster than $T_1$ and $T_2$ ($\sim 10^2/J$). This is consistent with depolarization driven by pair creation, rather than by pulse errors. Notably, $\ket{\uparrow}$ excitations emerge first at the edges, as they have fewer neighbors contributing to the Ising energy cost of spin flips.

To show that the depolarization is indeed due to pair creation, we extract the full counting statistics from the experimental snapshots. Fig.~\ref{fig_4}(c) shows $P_\text{FC}({N_\uparrow},t)$, the probability of observing $N_\uparrow$ spins in $\ket{\uparrow}$. Fig.~\ref{fig_4}(d) shows the measured parity $\langle \hat{\mathcal{P}} \rangle$ along with the total polarization $\langle \hat{S}^z \rangle$.  We find that over the rapid timescale on which $|\langle \hat{S}^z \rangle|$ decreases, $\langle \hat{\mathcal{P}} \rangle$ remains high, directly revealing the parity conservation in the anisotropic XYZ model. We attribute the parity decay to single particle depolarization and higher order Floquet terms that arise from local disorder caused by differential tweezer light shifts~\cite{Supplement}. We note that parity, being a global observable, decays $\sim N_{\text{mol}}$ times faster than the single particle $T_1$. We also observe that $\langle\hat{S}^z \rangle$ shows hints of an oscillation, which is a signature of coherent dynamics. The data agrees well with simulations of the ideal XYZ model, indicating that over the pair creation timescale, interaction disorder and decoherence play a minimal role.

To conclusively show coherent pair creation and also annihilation, we next identify a regime of interactions parameters where these processes can be cleanly observed. In the limit of $\Delta\gg1$ with nearest-neighbor interactions, the domain wall number $N_D$ is approximately conserved. Here, the Hilbert space is partitioned into domain wall sectors separated by an energy of $J \Delta/2$ (Fig. \ref{fig_4}(e)). An initial product state $\ket{\psi_D}=\ket{\downarrow\downarrow\uparrow\uparrow\uparrow\uparrow\uparrow\uparrow}$, containing one domain wall, is only coupled to states with $N_D=1$, meaning that the initial $\ket{\downarrow}$ domain can only change size but not break apart. In addition, parity conservation implies that only even domain sizes are allowed for $N_{\text{mol}}=8$. Thus, coherent pair creation and annihilation is mapped onto a quantum walk of the domain wall with a step size of two sites and a hopping amplitude of $t_{\rm{eff}}= \gamma J/2$. The domain wall position is directly accessible from our site-resolved spin readout.

Experimentally, we prepare the system in $\ket{\psi_D}$ and measure $\left\langle \hat{S}^z_i\right\rangle $ versus time for $\Delta=1.860\,(13)$ and $\gamma=0.290(3)$. We post-select for shots with no vacancies and even parity. Our site-resolved measurements (Fig. \ref{fig_4}(g)) indeed show growth and shrinkage of the $\ket{\downarrow}$ domain caused by coherent creation and annihilation of spin pairs at the domain boundary. The shrinkage corresponds to the domain wall reflecting off the boundary, a hallmark of coherent quantum walk dynamics (Fig.~\ref{fig_4}(f)). For further support, we also measure the microstate distribution in the $\hat{S}^z_i$ basis. As shown in Fig.~\ref{fig_4}(h), at short times, the system primarily explores the three states $\ket{\downarrow\downarrow\uparrow\uparrow\uparrow\uparrow\uparrow\uparrow}$, $\ket{\downarrow\downarrow\downarrow\downarrow\uparrow\uparrow\uparrow\uparrow}$, and $\ket{\downarrow\downarrow\downarrow\downarrow\downarrow\downarrow\uparrow\uparrow}$. Fig.~\ref{fig_4}(i) shows that the corresponding bitstring probabilities oscillate, directly establishing coherent pair creation and annihilation dynamics. Notably, these constitute the first microscopic observations of coherent pair creation and annihilation in XYZ-type spin models in atomic or molecular quantum simulators.

\section{Summary and Outlook}
In summary, we have realized tunable $1/r^3$ XXZ/XYZ spin chains in the novel platform of molecular tweezer arrays and microscopically probed a variety of coherent many-body spin dynamics. Notably, we have observed, for the first time, two-magnon bound states in $1/r^3$ XXZ models and coherent pair creation and annihilation dynamics in $1/r^3$ XYZ models. Our results establish molecular tweezer arrays as a new quantum simulation platform well-suited for exploring dipolar quantum spin models. While our work has focused on dynamics, it also opens the door to investigating low temperature states of these models and their magnetic phases. Looking ahead, larger molecular arrays achievable with technical improvements in molecular loading could allow investigation of dipolar spin models in 2D geometries, opening access to simulating novel phenomena such as geometric frustration and spin liquids~\cite{Savary2017SpinLiquid}. Much longer coherent many-body evolution times could also be achieved by reducing interaction disorder, for example by cooling the molecules to the motional ground state within a tweezer~\cite{Lu2024RSCCaF, Bao2024RSCCaF}. In addition to larger system sizes, new geometries, and longer coherent evolution times, higher spins can also be encoded in multiple molecular states and coherently controlled~\cite{Holland2025Erasure}. This feature opens the door to richer Hamiltonians such as spin-1 models, bosonic $t$-$J$ models~\cite{Lukas2024tJ}, and lattice gauge theories~\cite{Halimeh2024LGT}. Our work also opens the door to quantum-enhanced metrology with molecules, since interacting spin models can be used to generate metrologically useful entangled states~\cite{Pezze2018RMPMetrology,Perlin2020spinsqueeze,Bornet2023spinsqueeze}, which are particularly relevant for precision measurement experiments with polar molecules~\cite{Hudson2002YbFEDM,ACME2018,JILAEDM2023}.

\section{Acknowledgements}
We thank Dima Abanin, Sarang Gopalakrishnan, Hyunsoo Ha, Rhine Samajdar, and David Huse for fruitful discussions. We also acknowledge early experimental help from Samuel Li and thank Christopher Tong for a careful reading of the manuscript. This work was supported by the National Science Foundation under Grant No. 2207518 and Grant No. 2513401, and by the Air Force Office of Scientific Research under Grant No. FA9550-24-1-0140 and Grant No. FA9550-25-1-0092. C. L. W. acknowledges support from a Princeton Quantum Initiative Graduate Student Fellowship. X.-Y. C. acknowledges support from a Dicke Postdoctoral Fellowship. L. W. C. acknowledges support from the Alfred P. Sloan Foundation under Grant No. FG-2022-19104.

\newpage

\end{document}


\begin{center}
	{\Large Supplementary Materials for ``Probing Coherent Many-Body Spin Dynamics in a Molecular Tweezer Array Quantum Simulator''}\\[1em]
	
	Yukai Lu,$^{1,2}$ Connor M. Holland,$^1$ Callum L. Welsh,$^1$ Xing-Yan Chen,$^1$ and Lawrence W. Cheuk$^{1,*}$\\[0.5em]
	{\small
		$^1$Department of Physics, Princeton University, Princeton, NJ 08544, USA\\
		$^2$Department of Electrical and Computer Engineering, Princeton University, Princeton, NJ 08544, USA\\
		$^*$Correspondence to: lcheuk@princeton.edu
	}
\end{center}

\vspace{1em}

\section{Materials and Methods}

\subsection{Experimental Sequence}

\subsubsection{Cooling and Loading Molecules into Optical Tweezer Traps}
A detailed description of our apparatus, together with procedures for laser cooling CaF and loading molecules into optical tweezer traps, can be found in Refs.~\cite{Holland2023Bichro, Holland2023SE}. In brief, CaF molecules are generated in a cryogenic buffer-gas source, decelerated using chirped slowing~\cite{Truppe2017chirp}, and captured in a DC magneto-optical trap (MOT)~\cite{Truppe2017Mot}. We then compress the MOT by increasing the magnetic-field gradient and perform $\Lambda$-enhanced gray-molasses cooling in free space~\cite{Cheuk2018Lambda}. Molecules are further compressed in a blue-detuned magneto-optical trap~\cite{Li2024CaFBDM} and subsequently transferred into a 1064-nm optical dipole trap (ODT). 

We next move the ODT to the focal plane of the objective and load molecules into the optical tweezers using $\Lambda$ cooling~\cite{Anderegg2019Tweezer}. Light-assisted collisions ensure that each tweezer trap contains at most one molecule. Our 1D optical tweezer array consists of up to 37 traps formed by focusing linearly polarized 781-nm light through a microscope objective~\cite{Holland2023Bichro}. In short, a 2D acousto-optic deflector (AOD) driven by an arbitrary waveform generator (AWG) diffracts a single laser beam into multiple orders. These are focused through the objective (NA = 0.65) to a waist of $w_0 = 730$ nm. The array configuration can be controlled in real time through the AWG.  After initial loading, we use a 4-ms nondestructive $\Lambda$-imaging pulse to identify the occupied tweezers. The occupied sites are then rearranged into a uniformly spaced 1D array of the desired size~\cite{Holland2023SE}.

\begin{figure}[h!]
\includegraphics{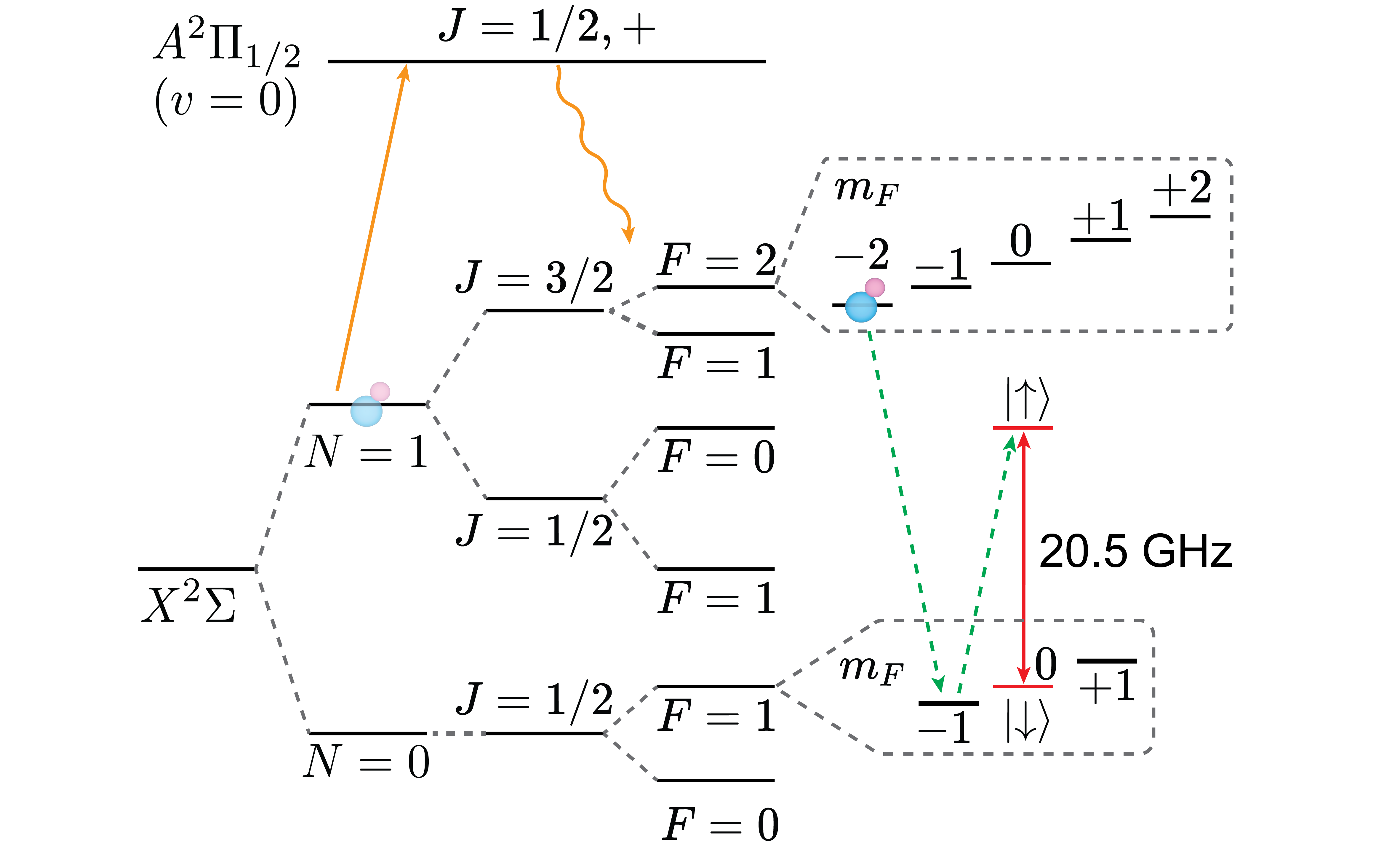}
\caption{\label{fig:S2_energy_level} Energy-level diagram of CaF molecules. The spin is encoded in the two rotational states $\ket{\downarrow}=\ket{N=0,J=1/2,F=1,m_F=0}$ and $\ket{\uparrow}=\ket{N=1,J=1/2,F=0,m_F=0}$. Molecules are prepared in $\ket{\uparrow}$ using optical pumping and microwave transfers, and the spin Hamiltonians are engineered by driving the transition between $\ket{\uparrow}$ and $\ket{\downarrow}$.}
\end{figure}

\subsubsection{Preparing the Polarized State $\ket{\uparrow}^{\otimes N_{\text{mol}}}$}
\label{sec:state_prep}
An energy-level diagram of the relevant rotational manifolds is shown in Fig.~\ref{fig:S2_energy_level}. The detailed measurement-enhanced state-preparation scheme can be found in Ref.~\cite{Holland2025Erasure}. In brief, we first optically pump molecules into $\ket{D^\prime}=\ket{N=1,F=2,m_F=-2}$. Subsequently, we ramp the magnetic field and optical tweezer depth into a configuration where the $\ket{D^\prime}-\ket{-}=\ket{N=0,F=1,m_F=-1}$ transition is minimally affected. A composite microwave $\pi$ pulse transfers molecules from $\ket{D^\prime}$ to $\ket{-}=\ket{N=0,F=1,m_F=-1}$. We then apply a rapid resonant imaging pulse that detects the population of $N=1$ molecules. This image identifies sites where molecules are not prepared in the desired state $\ket{-}$, and is used in post-selection to reject data. Next, we apply a composite microwave pulse to transfer molecules from $\ket{-}$ to $\ket{\uparrow}$. 

After preparing molecules in $\ket{\uparrow}$, the magnetic field is rotated to be perpendicular to the tweezer polarization for local state preparation and subsequent time evolution under $\hat{H}_{\text{XYZ}}$. We wait 100\,ms for the magnetic field to settle. Blackbody radiation can excite molecules into the $v=1$ vibrational manifold during this interval, causing state-preparation errors at the $\approx 3\%$ level. The overall state-preparation fidelity up to this point is $\approx 90\%$, with errors predominantly being vacant sites.

\subsubsection{Preparing Arbitrary Product States of $\ket{\uparrow}$ and $\ket{\downarrow}$}\label{sec:local_state_prep}
To initialize the system into arbitrary product states of $\ket{\uparrow}$ and $\ket{\downarrow}$, we make use of local ac Stark shifts in combination with global microwave pulses. Specifically, we begin by preparing all molecules in the fully polarized state $\ket{\uparrow}^{\otimes N_{\text{mol}}}$ in tweezers of depth $U_0=k_B\times130\,{\rm \mu K}$. To flip selected spins from $\ket{\uparrow}$ to $\ket{\downarrow}$, we keep the target tweezers at depth $U_0$ while increasing the depth of all other tweezers to $U_{\rm SS}=5U_0$. Due to the differential ac Stark shift of the $\ket{\uparrow}\leftrightarrow\ket{\downarrow}$ transition, the resonant microwave frequency of the target sites is shifted by 20\,kHz relative to all other sites. A subsequent composite microwave $\pi$-pulse (Knill pulse) resonant with the target molecules selectively transfers them to $\ket{\downarrow}$. Each pulse in the Knill composite sequence has a Gaussian envelope with total duration 166 $\mu$s, which minimizes off-resonant excitation. We measure a flip fidelity of $f_{\rm flip}=99.84(6)\%$ and an off-resonant excitation probability of $p_{\rm off-res}=0.6(2)\%$. 

\subsection{Floquet Hamiltonian Engineering}
\subsubsection{Floquet Pulse Sequences}
The native Hamiltonian of the system is
\begin{equation}\label{eq:native_hamiltonian}
\hat{H}_s = \sum_{i<j} \frac{J_{\rm ex}}{|j-i|^3}(\hat{S}_i^x\hat{S}_j^x + \hat{S}_i^y\hat{S}_j^y) + \sum_i h_i \hat{S}_i^z,
\end{equation}
where the first term describes the native dipolar spin-exchange interaction and the second term describes on-site disorder primarily due to differential light shifts. Throughout this supplement, we set $\hbar=1$. Our microwave pulses are global and are described by a control Hamiltonian
\begin{equation}\label{eq:control_hamiltonian}
\hat{H}_c(t) = \sum_i \Omega(t) [\cos\phi(t) \hat{S}_i^x + \sin \phi(t) \hat{S}_i^y].
\end{equation}
To control the amplitude and phase of the microwave pulses, we use a signal generator with IQ modulation driven by an AWG.  

\begin{figure}
\includegraphics{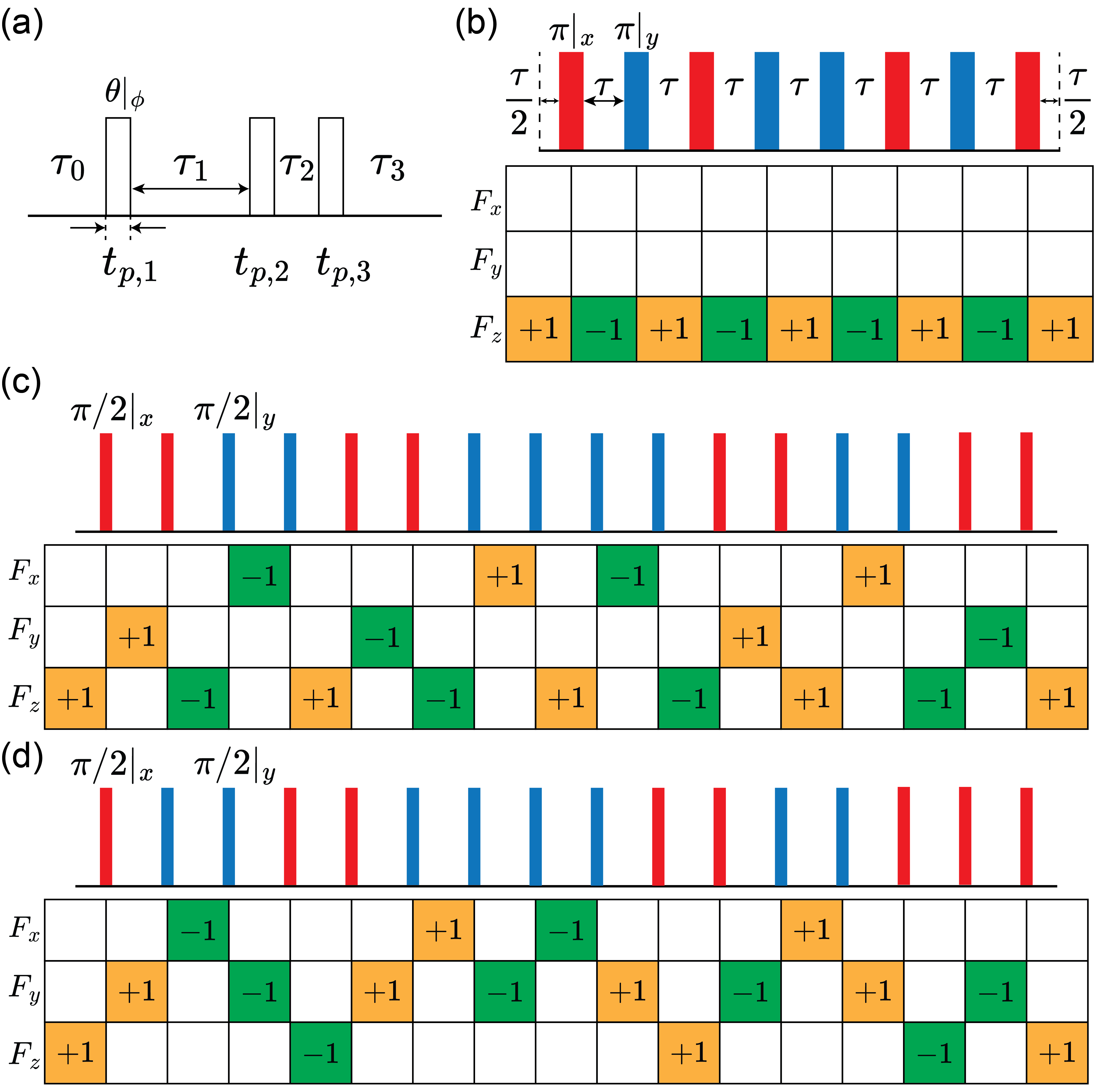}
\caption{\label{fig:S3_floquet} Diagram of the Floquet-engineering sequences. (a) A general sequence consists of pulses $\theta|_\phi$ of width $t_{w,k}$ separated by free-evolution intervals $\tau_k$. (b) XY8 dynamical-decoupling sequence. (c) XY8-XYZ Floquet-engineering sequence obtained by splitting each $\pi$ pulse in XY8 into two $\pi/2$ pulses. (d) Shifted XY8-XYZ sequence used to engineer XYZ models. Following the toggling-frame framework of Ref.~\cite{choi2020robust}, the shaded orange (green) boxes show which axis of the toggling frame is aligned (anti-aligned) with the lab $z$-basis.}
\end{figure}

\begin{table*}
\centering
\begin{ruledtabular}
\begin{tabular}{lcccccc}
$\Delta$ & $\gamma$ & Sequence name & $\tau_{\rm XY}\,(\mu{\rm s})$ & $\tau_{\rm XZ}\,(\mu{\rm s})$ & $\tau_{\rm YZ}\,(\mu{\rm s})$ & $T\,(\mu{\rm s})$ \\
\hline
0.012(11) & 0 & XY8 & 400 & N.A. & N.A. & 3226.4 \\
0.361(8) & 0 & XY8-XYZ & 66.7 & 24.7 & 24.7 & 784 \\
0.686(10) & 0 & XY8-XYZ & 46.7 & 46.7 & 46.7 & 800 \\
1.000(12) & 0 & XY8-XYZ & 21.7 & 46.7 & 46.7 & 600 \\
1.329(11) & 0 & XY8-XYZ & 8.7 & 44.7 & 44.7 & 480 \\
1.741(19) & 0 & XY8-XYZ & 2 & 91.4 & 91.4 & 800 \\
1.033(11) & 0.466(5) & shifted XY8-XYZ & 61.7 & 59.05 & 2 & 780 \\
1.860(13) & 0.290(3) & shifted XY8-XYZ & 2 & 59.05 & 61.7 & 780 \\
\end{tabular}
\end{ruledtabular}
\caption{\label{tab:S1_floquet_sequences} Floquet-engineering pulse sequences used to realize the XXZ and XYZ Hamiltonians explored in this work. Here $\tau_{\rm XY}$, $\tau_{\rm XZ}$, and $\tau_{\rm YZ}$ denote the free-evolution times spent in the corresponding toggling frames within one Floquet cycle, and $T$ is the Floquet period. These free-evolution times are measured between the edges of adjacent pulses. For all sequences, the pulse width is $t_p=3.3\,\mu{\rm s}$. For the shifted XY8-XYZ sequences, our minimum-pulse-spacing constraint makes the 8th free-evolution interval $\tau_8=\tau_{\rm XZ}-2.65\,\mu{\rm s}=56.4\,\mu{\rm s}$, while the 16th free-evolution interval is $\tau_{16}=\tau_{\rm XZ}+2.65\,\mu{\rm s}=61.7\,\mu{\rm s}$.}
\end{table*}

Our Floquet sequence consists of pulses $\{\hat{P}_k\}$ separated by free-evolution intervals $\{\tau_k\}$, as shown schematically in Fig.~\ref{fig:S3_floquet}(a). The pulses are typically short compared to $\tau_k$. Effectively, each pulse rotates the native Hamiltonian in the interaction picture (toggling frame). During the free-evolution interval $\tau_k$ after the $k$th pulse, the system evolves according to the Hamiltonian
\begin{equation}
\hat{\tilde{H}}_k = (\hat{P}_k\cdots \hat{P}_1\hat{P}_0)^\dagger \hat{H}_s(\hat{P}_k\cdots \hat{P}_1\hat{P}_0),
\end{equation}
where $\hat{P}_0=\mathbb I$. 

Since $J_{\rm ex}T\ll 1$ in our experiment, where $T$ is the sequence period, we can expand the effective Hamiltonian in powers of $J_{\rm ex}T\ll 1$. We employ sequences that are symmetric in time (up to a shift in the cycle), which implies that the effective Hamiltonian is given by the weighted average of $\tilde H_k$ over one Floquet cycle, with higher-order corrections entering at $(J_{\text{ex}}T)^2 \sim \mathcal{O}(10^{-2})$.

Our sequences are designed to cancel static on-site disorder and thereby enable long coherent evolution times compared to $1/J_{\rm ex}$. To analyze and design these sequences, we use the toggling-frame framework of Ref.~\cite{choi2020robust}, in which the control is represented by the toggled spin frame matrix
\begin{equation}
F_{\alpha,k}=2\,\mathrm{Tr}[\hat{S}^\alpha \hat{\tilde{S}}_k^z],
\end{equation}
with $\hat{\tilde{S}}_k^z$ the transformed spin operator in the $k$-th interval.

To realize the native XY model, we use the standard XY8 dynamical decoupling sequence, which suppresses static and slow-varying on-site disorder. The XY8 sequence consists of eight $\pi$ pulses along the $\hat{x}$ or $\hat{y}$ directions (Fig.~\ref{fig:S3_floquet}(b)). To minimize off-resonant excitation to other nearby hyperfine states, we employ square-top pulses with soft edges with a total width $t_p=3.3\,\mu$s, separated by free-evolution intervals $\tau=400\,\mu$s. 

To engineer XYZ spin models, we use $\pi/2$ pulses to toggle among different spin frames. The Hamiltonian in each toggling frame $\hat{\tilde{H}}_k$ takes three distinct forms: $\hat{S}_i^x \hat{S}_j^x + \hat{S}_i^y \hat{S}_j^y$ (XY), $\hat{S}_i^x\hat{S}_j^x + \hat{S}_i^z\hat{S}_j^z$ (XZ), and $\hat{S}_i^y\hat{S}_j^y + \hat{S}_i^z \hat{S}_j^z$ (YZ). Building on the XY8 sequence, we construct the XY8-XYZ sequence by splitting each $\pi$ pulse in the XY8 sequence into two $\pi/2$ pulses along the same rotation axis. This yields 16 pulses per cycle (Fig.~\ref{fig:S3_floquet}(c)). We empirically find that this type of sequence gives the longest measured $T_1$ and $T_2$. The free-evolution times spent in the XY, XZ, and YZ frames, denoted $\tau_{\rm XY}$, $\tau_{\rm XZ}$, and $\tau_{\rm YZ}$, control the engineered couplings. For XYZ models, in order to achieve maximum strength in either $\gamma J$ or $J \Delta$, we use a shifted variant in which the first $\pi/2$ pulse is moved to the end of the sequence (Fig.~\ref{fig:S3_floquet}(d)). This sequence maintains the time-symmetric structure and also has higher-order Floquet terms entering at order $(J_{\text{ex}}T)^2 \sim \mathcal{O}(10^{-2})$.

The full sequence parameters for all sequences used in this work are provided in Table~\ref{tab:S1_floquet_sequences}.

\subsubsection{Single-Particle Characterization of Floquet Sequence Performance}\label{sec:S_single_particle}
We first characterize the Floquet sequences at the single-particle level, where both coherent and incoherent errors need to be considered. In the absence of errors, the single-spin propagator over one Floquet cycle is the identity, $\hat U_c(T)=\mathbb I$. Coherent pulse errors can be described by an effective magnetic field $\vec{B}_{\text{eff}}=\{B_{\text{eff}}^x, B_{\text{eff}}^y, B_{\text{eff}}^z\}$ with Hamiltonian 
\begin{equation}
	\hat{H}_p=\vec{B}_{\text{eff}}\cdot \vec{S}
\end{equation}
resulting in the time-evolution propagator
\begin{equation}
	\hat{U}_c(T)  = e^{-i\hat{H}_p T}.
\end{equation}

We determine $\vec B_{\text{eff}}$ using three measurements: 

(i) prepare $\ket{\uparrow}$, apply a variable number of Floquet cycles, then apply a pulse $\frac{\pi}{2}|_0$ followed by an analysis pulse $\frac{\pi}{2}|_\phi$; 

(ii) prepare $\ket{\uparrow}$, apply a variable number of Floquet cycles, then apply a pulse $\frac{\pi}{2}|_{\pi/2}$ followed by an analysis pulse $\frac{\pi}{2}|_\phi$; 

(iii) perform a standard Ramsey sequence, $\frac{\pi}{2}|_0$ -- variable Floquet cycles -- $\frac{\pi}{2}|_\phi$. 

\begin{figure}[h!]
\includegraphics[width=\textwidth]{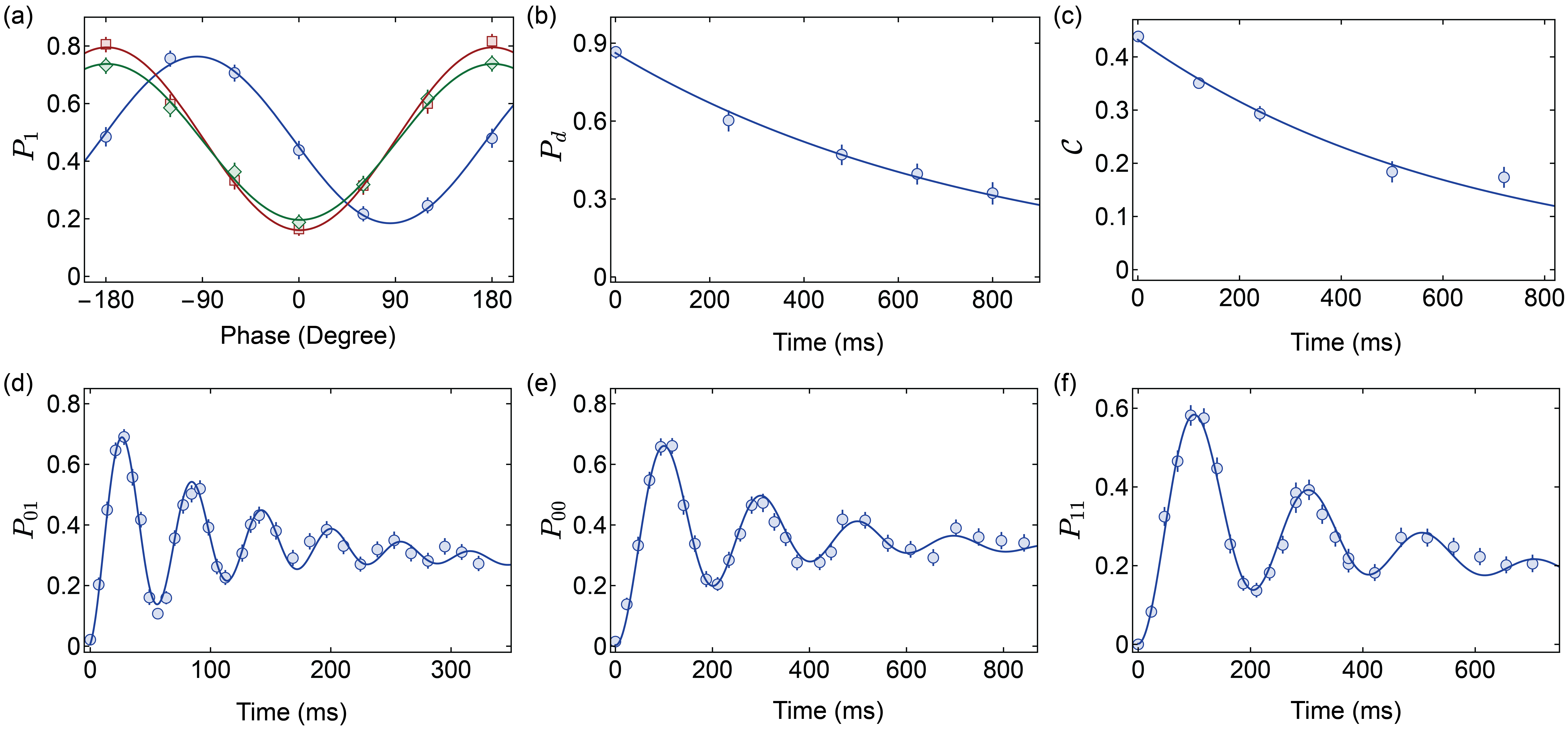}
\caption{\label{fig:S4_seqchar} Characterization of the Floquet-engineering sequences. The top panels (a-c) show single-particle characterization of the sequence used to engineer the XXZ model with $\Delta=0.686(10)$. (a) Ramsey fringes after 300 Floquet cycles (240 ms). Blue circles, red squares, and green diamonds correspond to measurements (i), (ii), and (iii) described in the text, indicating negligible $\vec B_{\rm eff}$. (b) $P_d=P_\uparrow-P_\downarrow$ as a function of time, yielding $T_1=0.80(4)$ s. (c) Decay of the coherence $\mathcal C$, yielding $T_2=0.64(5)$ s. The bottom panels (d-f) show two-particle characterization of the sequence used to engineer the XYZ model with $\Delta=1.860(13)$ and $\gamma=0.290(3)$. They correspond to measurements (1), (2), and (3) described in the text.}
\end{figure}

When $B_{\text{eff},x}=0$, (i) gives a Ramsey fringe with phase of $90^\circ$ while (ii) and (iii) give fringes with phase of $0^\circ$ when $B_{\text{eff}, y}=0$ and $B_{\text{eff}, z}=0$. With these three measurements, we verify $\vec{B}_{\text{eff}}\approx 0$ for all Floquet sequences (Fig. \ref{fig:S4_seqchar}(a)).

To characterize incoherent errors arising from control imperfections and on-site disorder, we measure the depolarization time $T_1$ and the decoherence time $T_2$. To measure $T_1$, we prepare molecules in $\ket{\uparrow}$, apply varying numbers of Floquet cycles, and measure $P_\uparrow$ and $P_\downarrow$, the populations in $\ket{\uparrow}$ and $\ket{\downarrow}$. We fit the population difference $P_d=P_\uparrow-P_\downarrow$ to $P_d= P_d(0) \exp(-t/T_1)$ to obtain $T_1$ (Fig.~\ref{fig:S4_seqchar}(b)). To measure $T_2$, we insert Floquet cycles into the standard Ramsey sequence and extract the coherence $\mathcal C$ as a function of time (Fig.~\ref{fig:S4_seqchar}(c)). The lifetime of the $\{\ket{\uparrow},\ket{\downarrow}\}$ manifold is independently measured to be $\approx 3.5$\,s, which is limited by blackbody radiation that excites molecules into vibrationally excited $v=1$ states~\cite{Holland2025Erasure}. The reported values of $T_1$ and $T_2$ in the main text are corrected for this background loss.

\subsubsection{Calibration of Effective Hamiltonian Parameters Using Two-Particle Dynamics}\label{sec:S_two_particle}
We calibrate the effective Hamiltonian $H_{\text{XYZ}}$ using two-molecule dynamics [Fig.~\ref{fig:S4_seqchar}(d-f)]. We perform three measurements by preparing molecules in three different product states. The subsequent dynamics allow the combinations $J_x+J_y$, $J_x-J_y$, and $J_x-J_z$ to be calibrated. In detail:

\begin{enumerate}
\item Starting from $\ket{\psi(t=0)}=\ket{\uparrow\downarrow}$, the molecules undergo coherent spin exchange, with
\begin{equation}
\ket{\psi(t)}=\cos\left(\frac{J_x+J_y}{4}t\right)\ket{\uparrow\downarrow}-i\sin\left(\frac{J_x+J_y}{4}t\right)\ket{\downarrow\uparrow}.
\end{equation}

\item Starting from $\ket{\psi(t=0)}=\ket{\uparrow\uparrow}$ (even-parity subspace), the two molecules evolve as
\begin{equation}
\ket{\psi(t)}=\cos\left(\frac{J_x-J_y}{4}t\right)\ket{\uparrow\uparrow}-i\sin\left(\frac{J_x-J_y}{4}t\right)\ket{\downarrow\downarrow}.
\end{equation}

\item Starting from $\ket{\psi(t=0)}=\ket{\uparrow\uparrow}$, we apply a Ramsey sequence with a variable number of Floquet cycles inserted between first and last $\frac{\pi}{2}|_0$ Ramsey pulses. This gives
\begin{equation}
\ket{\psi(t)}=\sin\left(\frac{J_x-J_z}{4}t\right)\ket{\uparrow\uparrow}+i\cos\left(\frac{J_x-J_z}{4}t\right)\ket{\downarrow\downarrow}.
\end{equation}
We fit the data to exponentially damped sinusoids. The extracted oscillation frequencies are then used to determine $J_x$, $J_y$, and $J_z$.
\end{enumerate}

\subsection{XXZ Experimental Sequence and Post-Selection Procedure}
\subsubsection{Detailed XXZ Experimental Sequence}
Our experimental sequence for exploring the XXZ models proceeds as follows. After stochastic loading into optical tweezers with spacing $4.20(6)\,\mu$m, we perform a 4-ms nondestructive $\Lambda$-imaging pulse to identify occupied tweezers. The occupied tweezers are then rearranged into a defect-free array of $N_{\rm mol}$ sites, followed by a second nondestructive $\Lambda$-imaging pulse for verification.

We then prepare all molecules in $\ket{\uparrow}$ using the procedure described in Sec.~\ref{sec:state_prep}. Next, we perform local state preparation as described in Sec.~\ref{sec:local_state_prep} to initialize either a single or two spin excitations. The tweezers are subsequently moved to the interacting configuration with spacing $2.1(3)\,\mu$m in 2\,ms, a variable number of Floquet cycles is applied, and the tweezers are returned to the loading configuration. A composite $\pi$ pulse is applied to swap $\ket{\uparrow}$ and $\ket{\downarrow}$ with fidelity $99.7(1)\%$. Finally, we use 30\,ms destructive imaging to detect molecules in $\ket{\uparrow}$ with fidelity $\sim 99.0\%$.

\subsubsection{XXZ Post-Selection Procedure}
To reject errors arising from imperfect state preparation and depolarization, we post-select on data according to several criteria. We define the status of each tweezer as $O_i^\alpha$, where $i$ is the tweezer index and $\alpha$ labels the image. Here, $O=1$ ($0$) denotes a bright (dark) tweezer following a fluorescence image. For the XXZ data presented in the main text, three images are taken each run: (1) a non-destructive verification image after initial tweezer rearrangement with status $O_i^v$, (2) a state-preparation error detection image with status $O_i^e$, and (3) a final destructive image that measures the $\ket{\uparrow}$ occupation of each tweezer with status $O_i^d$.

Specifically, we post-select for experimental shots satisfying three criteria:
\begin{enumerate}
\item a defect-free array is observed in the non-destructive verification image after initial rearrangement, $\sum_i O_i^v = N_{\rm mol}$;
\item no state-preparation errors in the error-detection image, $\sum_i O_i^e = 0$;
\item the correct number of bright sites is observed in the final image, $\sum_i O_i^d = N_{\rm mol}-N_s$, where $N_s$ is the number of initialized $\ket{\uparrow}$.
\end{enumerate}
The third criterion specifically makes use of  $\hat{S}_z$ conservation in the XXZ model, the same strategy employed in Ref.~\cite{Google2022PhotonBoundState}.

We note that the final detection image for the XXZ data does not distinguish an empty site from a molecule in $\ket{\uparrow}$. Since the state preparation infidelity, after error detection, arises from vacant sites, if the initial location of $\ket{\uparrow}$ aligns with an error site, the data still passes post-selection. This means that, for example, for the quantum walk data, there will be a small background where the spin $\ket{\uparrow}$ appears stuck in its initial site. This error is estimated to be $\approx 10\%$. 

\subsection{XYZ Experimental Sequence and Post-Selection Procedure}
\subsubsection{Detailed XYZ Experimental Sequence}
The experimental sequence used to study XYZ models follows the same overall structure as for the XXZ measurements, with small differences in state preparation and final detection.

After stochastic loading and rearrangement into a defect-free array, we prepare molecules in $\ket{\downarrow}$ using a modified sequence that allows us to reject additional leakage errors caused by blackbody radiation~\cite{Holland2025Erasure} during the hold time used for the magnetic fields to settle. Specifically, we first optically pump molecules into $\ket{N=1,J=3/2,F=2,m_F=-2}$, then use a composite $\pi$ pulse to transfer them to $\ket{N=0,J=1/2,F=1,m_F=-1}$. This is followed by a second composite $\pi$ pulse to transfer them to $\ket{\uparrow}$. We then transfer the molecules to $\ket{\downarrow}$ at the optimal magnetic field. After waiting 100\,ms for magnetic-field stabilization, we apply a rapid image pulse and detect molecules in the $N=1$ manifold resulting from imperfect state preparation and blackbody radiation. Here, molecules in $\ket{\downarrow}$ that undergo blackbody leakage end in the $v=1, N=1$ manifold, which is then repumped into the $v=0, N=1$ imaging manifold during the state preparation error detection image.

For parity-dynamics measurements, we initialize all molecules in $\ket{\downarrow}$, move the tweezers to the interacting configuration with spacing $2.1\,\mu$m in 2 ms, apply a variable number of Floquet cycles, return the tweezers to the loading configuration, and then perform spin-resolved imaging. For domain-wall measurements, we use the local state-preparation procedure described in Sec.~\ref{sec:local_state_prep} to prepare the initial state $\ket{\downarrow\downarrow\uparrow\uparrow\uparrow\uparrow\uparrow\uparrow}$ and then apply the same interaction and detection sequence.

\subsubsection{Full Spin-Resolved Readout}
To detect both $\ket{\uparrow}$ and $\ket{\downarrow}$, we use a two-step spin-resolved imaging sequence that uses the multiple internal states of the CaF molecule. We first increase the tweezer depth to $k_B\times1\,\rm mK$ and optically pump molecules from $\ket{\uparrow}$ to the $N=3$ rotational manifold in 0.2\,ms using resonant light addressing the $\ket{X,v=0,N=1}-\ket{B,v=0,N=2,J=3/2}$ transition. Next, we decrease the tweezer depth to $U=U_0$ and apply a composite $\pi$ pulse to transfer molecules from $\ket{\downarrow}$ to $\ket{\uparrow}$. Detection of this population is performed using a combination of 4\,ms $\Lambda$ imaging and 7\,ms rapid resonant imaging~\cite{Holland2025Erasure} after turning off the magnetic field and ramping the tweezers to $k_B\times1.3$ mK. During this imaging sequence, rotational repumping light addressing the transition $\ket{X,v=0,N=3}\to\ket{B,v=0,N=2,J=5/2}$ is intentionally turned off so that molecules shelved in $N=3$ are not imaged. This imaging pulse is largely destructive for $\ket{\downarrow}$. To detect molecules in $\ket{\uparrow}$, we next employ standard 30\,ms destructive $\Lambda$-imaging with the rotational repumping light restored, which returns shelved molecules from $N=3$ to the $N=1$ imaging manifold.

For each tweezer, we bin the two spin readout images and sum the camera counts to construct separate histograms for the two images. After thresholding, each site is assigned one of four possible outcomes: 00, 10, 01, and 11, corresponding to empty ($\ket{\emptyset}$), a molecule in $\ket{\downarrow}$, a molecule in $\ket{\uparrow}$, and the situation where both spin images are bright. Since the molecule can only be projected to either $\ket{\downarrow}$ or $\ket{\uparrow}$, the last outcome 11 arises from $\ket{\downarrow}$ molecules that survived the first image and also classification errors in the two images. In the data analysis, the outcomes 10 and 11 are both classified as measured $\ket{\downarrow}$. The spin-resolved imaging can be characterized by a $3\times 3$ process matrix $M$, defined through
\begin{equation}
	\mathbf{p}_{\rm meas}=M\,\mathbf{p}_{\rm true},
\end{equation}
which relates the true probabilities to the measured binarized outcomes. Here $M_{\mu\nu}$ denotes the probability to identify a site in measured state $\mu$ given that its true state is $\nu$. Explicitly, using separate calibration data, we obtain
\begin{equation}
	M=
	\begin{pmatrix}
		0.956(23) & 0.0187(22) & 0.00054(18) \\
		0.0086(15) & 0.934(23) & 0.00029(13) \\
		0.035(7) & 0.048(7) & 0.999(24)
	\end{pmatrix},
\end{equation}
where the basis is ordered as $(\ket{\uparrow},\,\ket{\downarrow},\,\ket{\emptyset})$. Each column sums to unity. The diagonal elements give the correct identification probabilities for $\ket{\uparrow}$, $\ket{\downarrow}$, and an empty site, while the off-diagonal elements quantify spin misidentification and loss.

\subsubsection{XYZ Post-Selection Procedure}
For investigations of parity dynamics in the XYZ model, we retain only experimental shots satisfying three criteria:
\begin{enumerate}
\item a defect-free array is observed in the verification image after rearrangement, $\sum_i O_i^v = N_{\rm mol}$;
\item no state-preparation errors are detected in the error-detection image, $\sum_i O_i^e = 0$;
\item the intended number of molecules is observed in the spin-resolved images, $\sum_i O_i^d = N_{\rm mol}$.
\end{enumerate}
We define $O_i^d=1$ when site $i$ is classified as occupied ($\ket{\uparrow}$ or $\ket{\downarrow}$) and $O_i^d=0$ when it is classified as empty. For pair creation and annihilation data, we impose one additional post-selection criterion using the parity symmetry of the XYZ model: the parity of the number of $\ket{\downarrow}$ in the final spin-resolved images must match that of the initial state. That is, it must be even for the data presented in the main text.

\section{Supplementary Text}

\subsection{Effective Hamiltonian under Floquet Engineering}\label{sec:S_text1}
\subsubsection{Effective Hamiltonian Theory}
The evolution of a periodically driven system can be described by an effective Hamiltonian $\hat{H}_{F}$. We follow the approach of Ref.~\cite{Eckardt2015floquet}, in which $\hat{H}_{F}$ is chosen to be independent of the starting phase of the periodic Hamiltonian. This corresponds to the gauge-invariant Hamiltonian that governs the long-time dynamics independently of the micromotion. In this framework, $\hat{H}_{F}$ can be obtained systematically from a high-frequency expansion,
\begin{equation}
	\hat{H}_{F}= \sum_i \hat{H}_F^{(i)},
\end{equation}
with the lowest-order terms given by
\begin{eqnarray}
	\hat{H}_{F}^{(0)}&=&0, \nonumber\\
	\hat{H}_{F}^{(1)}&=&\frac{1}{T}\int_0^T dt\, \hat{H}(t), \nonumber\\
	\hat{H}_{F}^{(2)}&=&\sum_{m\neq0} \frac{T}{2\pi \hbar m}\,\hat{H}_m\hat{H}_{-m}, \nonumber\\
	\hat{H}_m &=& \frac{1}{T} \int_0^T dt\, e^{-i m 2\pi t/T} \hat{H}(t).
\end{eqnarray}

For a periodic Hamiltonian that is symmetric about a point in its cycle, one can choose the time origin at that symmetry point so that $\hat H(t)=\hat H(-t)$. In that case, the Fourier components satisfy
\begin{eqnarray}
	\hat{H}_m &=& \frac{1}{T} \int_{-T/2}^{T/2} dt\, e^{-i m 2\pi t/T} \hat{H}(t) \nonumber\\
	&=& \frac{2}{T} \int_{0}^{T/2} dt\, \cos(m 2\pi t/T)\hat{H}(t),
\end{eqnarray}
which implies
\begin{equation}
	\hat H_m=\hat H_{-m}.
\end{equation}
It then follows that the second-order term vanishes:
\begin{eqnarray}
	\hat{H}_{F}^{(2)}
	&=&
	\sum_{m\neq0} \frac{T}{2\pi \hbar m}\,\hat{H}_m\hat{H}_{-m}
	\nonumber\\
	&=&
	\sum_{m>0} \frac{T}{2\pi \hbar m}\,\hat{H}_m\hat{H}_{-m}
	+
	\sum_{m<0} \frac{T}{2\pi \hbar m}\,\hat{H}_m\hat{H}_{-m}
	\nonumber\\
	&=&0,
\end{eqnarray}
since the contributions from $m$ and $-m$ cancel pairwise.

All sequences used in this work (Table~\ref{tab:S1_floquet_sequences} and Fig.~\ref{fig:S3_floquet}) are time-symmetric. This implies that $\hat{H}_{F}^{(2)}$ is zero as long as all disorder terms are static. The leading order correction arises at the order of $(J_{\text{ex}} T)^2 \sim \mathcal{O}(10^{-2})$.

\subsubsection{Effective Hamiltonian and Accessible Parameter Region.}
Up to first order in $J_{\text{ex}}T$, the effective Hamiltonian is
\begin{equation}
H_{F}=\sum_{i<j}\frac{1}{|j-i|^3}\left(J_xS_i^xS_j^x+J_yS_i^yS_j^y+J_zS_i^zS_j^z\right),
\end{equation}
where the engineered couplings $J_x$, $J_y$, and $J_z$ are set by the total time spent in segments containing the corresponding bilinear term,
\begin{equation}
J_\alpha = \frac{J_{\rm ex}}{T}\sum_{k\in\mathcal U^\alpha}(\tau_k+t_p).
\end{equation}
Here $\mathcal U^\alpha$ denotes the set of free-evolution segments for which $\tilde H_k$ contains $S_i^\alpha S_j^\alpha$, and the finite pulse duration $t_p$ is included as an effective contribution to the total evolution time~\cite{choi2020robust}. Equivalently,
\begin{equation}
H_{F}=\sum_{i<j}\frac{J}{|j-i|^3}\left[\frac{1}{2}(S_i^+S_j^-+S_i^-S_j^+)
+\frac{\gamma}{2}(S_i^+S_j^+ + S_i^-S_j^-)+\Delta S_i^zS_j^z\right],
\end{equation}
with
\begin{equation}\label{eq:S_J_gamma_delta}
J = (J_x+J_y)/2,\qquad \Delta = 2J_z/(J_x+J_y),\qquad \gamma = (J_x-J_y)/(J_x+J_y).
\end{equation}
Because Floquet engineering redistributes the native interactions among different frames while preserving their sign, the couplings satisfy
\begin{equation}
J_x+J_y+J_z=2J_{\rm ex}, \qquad 0\le J_\alpha \le J_{\rm ex}.
\end{equation}
In the $(J,\gamma,\Delta)$ parametrization this becomes
\begin{equation}
J\left(1+\frac{\Delta}{2}\right)=J_{\rm ex},
\end{equation}
and, using Eq.~\eqref{eq:S_J_gamma_delta},
\begin{equation}
\begin{aligned}
J_x = \frac{2J_{\rm ex}(1+\gamma)}{2+\Delta}, \qquad J_y = \frac{2J_{\rm ex}(1-\gamma)}{2+\Delta}, \qquad J_z = \frac{2J_{\rm ex}\Delta}{2+\Delta}.
\end{aligned}
\end{equation}
Requiring $0\le J_\alpha\le J_{\rm ex}$ yields the accessible region
\begin{equation}
0\le \Delta \le 2, \qquad |\gamma|\le \frac{\Delta}{2}.
\end{equation}

\subsection{Effect of Finite Motional Temperature}\label{sec:Supp_ThermalMotion}
Due to finite motional temperature of the molecules within each tweezer, the intermolecular distances, and hence the interaction strengths, are modified. To lowest order, this modifies the interaction strength, as observed in experiments with two molecules~\cite{Holland2023SE, Bao2023SE}. If one assumes that molecules are in eigenstates of motion, then the experimental data averages over a distribution of $J$. Here, since we expect the effect to be largely due to thermal occupation in axial modes of the tweezer, which is far from the quantum regime, we model the thermal motion classically and derive the corresponding effective spin Hamiltonian.

\subsubsection{Motion Modulation}\label{MotionMod}
We model the motion of each molecule relative to the tweezer center as harmonic oscillation,
\begin{equation}
	\vec r_i = (q_{i,x}\cos\omega_r t + v_{i,x}\sin\omega_r t)\hat x' + (q_{i,y}\cos\omega_r t + v_{i,y}\sin\omega_r t)\hat y' 
	+ (q_{i,z}\cos\omega_z t + v_{i,z}\sin\omega_z t)\hat z',
\end{equation}
where $\omega_r$ and $\omega_z$ are the radial and axial trap frequencies, respectively. The coefficients $q_{i,\beta}$ and $v_{i,\beta}$ are sampled from Maxwell-Boltzmann distributions set by the measured temperatures. We assume $\omega_x=\omega_y=\omega_r$. The array direction is along $\hat x'$ while the quantization axis is along $\hat y'$.

Assuming all tweezers have the same radial and axial trapping frequencies, the Hamiltonian can be written as
\begin{equation}
	H_{\rm XY} = \sum_{i<j}\frac{J[1-3\cos^2\theta_{ji}' ]}{|(j-i)\hat x' + \vec r_j(t)-\vec r_i(t)|^3}(S_i^xS_j^x+S_i^yS_j^y),
\end{equation}
where
\begin{equation}
	\cos\theta_{ij}' = \frac{[(j-i)\hat x' + \vec r_j(t)-\vec r_i(t)]\cdot\hat y'}{|(j-i)\hat x' + \vec r_j(t)-\vec r_i(t)|}.
\end{equation}
Since the oscillation amplitude is small compared to the intermolecular distance, we Taylor expand the geometric factor,
\begin{equation}
	C_{ji}=\frac{1-3\cos^2\theta_{ji}'}{|(j-i)\hat x' + \vec r_j(t)-\vec r_i(t)|^3}\approx \frac{1}{|j-i|^3}\eta_{ji},
\end{equation}
with
\begin{equation}
	\eta_{ji}=1-3x_{ji}+6x_{ji}^2-10x_{ji}^3+15x_{ji}^4-\frac{9}{2}y_{ji}^2-\frac{3}{2}z_{ji}^2+\frac{45}{2}x_{ji}y_{ji}^2+\frac{15}{2}x_{ji}z_{ji}^2+\frac{75}{8}y_{ji}^4+\frac{15}{8}z_{ji}^4-\frac{135}{2}x_{ji}^2y_{ji}^2-\frac{45}{2}x_{ji}^2z_{ji}^2+\frac{45}{4}y_{ji}^2z_{ji}^2,
\end{equation}
where we define
\begin{equation}
	\beta_{ji}=\frac{1}{|j-i|}[q_{ji,\beta}\cos(\omega_\beta t)+v_{ji,\beta}\sin(\omega_\beta t)],
\end{equation}
with $q_{ji,\beta}=q_{j,\beta}-q_{i,\beta}$ and $v_{ji,\beta}=v_{j,\beta}-v_{i,\beta}$ for $\beta=x,y,z$.

Considering dynamics on a timescale $T_D$ satisfying $\omega_r T_D,\omega_z T_D\gg 1$ and $JT_D\ll 1$, we perform a Magnus expansion and retain the lowest-order term. The resulting effective Hamiltonian is
\begin{equation}
	H_{\rm XY,{\rm eff}}\approx \sum_{i<j}\frac{J}{|j-i|^3}(S_i^xS_j^x+S_i^yS_j^y)\bar\eta_{ji},
\end{equation}
where
\begin{equation}
	\bar\eta_{ji}=\frac{1}{T_D}\int_0^{T_D}\eta_{ji}\,dt.
\end{equation}
Since $\omega_rT_D,\omega_zT_D\gg 1$, this yields
\begin{equation}
	\bar\eta_{ji}=1+\frac{3\rho_{ji,x}^2}{(j-i)^2}-\frac{9\rho_{ji,y}^2+3\rho_{ji,z}^2}{4(j-i)^2}+\frac{45\rho_{ji,x}^4}{8(j-i)^4}+\frac{225\rho_{ji,y}^4+45\rho_{ji,z}^4}{64(j-i)^4}-\frac{135\rho_{ji,x}^2\rho_{ji,y}^2}{8(j-i)^4}-\frac{45\rho_{ji,x}^2\rho_{ji,z}^2}{8(j-i)^4}+\frac{45\rho_{ji,y}^2\rho_{ji,z}^2}{16(j-i)^4},
	\label{eq:motion_factor}
\end{equation}
where $\rho_{ji,\beta}^2=q_{ji,\beta}^2+v_{ji,\beta}^2$. The net effect is a shot-to-shot disorder in the dipolar couplings $J$.

\subsubsection{Floquet effective Hamiltonian in the presence of motion}
We now include both thermal motion and the microwave control sequence. The full time-dependent Hamiltonian is
\begin{equation}
	H(t)\approx \sum_{i<j}\frac{J}{|j-i|^3}\eta_{ji}(t)(S_i^xS_j^x+S_i^yS_j^y)+H_c(t).
\end{equation}
Considering the evolution at integer multiples of the Floquet period $T$, the effective Hamiltonian then takes the form
\begin{equation}
	H_{\rm eff}=\sum_{i<j}\frac{J}{|j-i|^3}(\bar\eta_{ji}^xS_i^xS_j^x+\bar\eta_{ji}^yS_i^yS_j^y+\bar\eta_{ji}^zS_i^zS_j^z),
	\label{eq:thermalHam}
\end{equation}
where
\begin{equation}\label{eq:thermal_eff_prefactor}
	\bar\eta_{ji}^\alpha=\frac{1}{T}\sum_{k\in\mathcal U_\alpha}\int_{T_{k-1}}^{T_k}\eta_{ji}\,dt.
\end{equation}
Here $T_k=\sum_{m\le k}\tau_m$ and $T_{-1}=0$. When $\omega_\alpha\tau_k\gg 1$, the segment average reduces to the time-averaged coupling discussed above, giving
\begin{equation}
	\bar\eta_{ji}^\alpha=\bar\eta_{ji}\frac{1}{T}\sum_{k\in\mathcal U_\alpha}(\tau_k+t_p).
\end{equation}
For some sequences, however, $\omega_\alpha\tau_k\sim 1$ and the relevant segment average must be evaluated explicitly. In that case,
\begin{equation}
	\begin{aligned}
		\frac{1}{T}\int_{T_{k-1}}^{T_k} [a\cos(\omega t)+b\sin(\omega t)]dt
		=\frac{\tau_k}{T}\frac{a\sin\phi_k-a\sin\phi_{k-1}+b\cos\phi_{k-1}-b\cos\phi_k}{\omega\tau_k},
	\end{aligned}
\end{equation}
where $\phi_k=\omega T_k$.

Although some free-evolution intervals in the Floquet engineering sequences are not strictly long compared with the motional period, Eq.~\eqref{eq:motion_factor} remains a good approximation for our experimental parameters. In our system, $J\approx 2\pi\times 33$ Hz, while the smallest trapping frequency is the axial one, $\omega_{\rm min}\sim 2\pi\times 13.3$ kHz. First, the trapping frequencies $\omega_\beta$ are incommensurate with the inverse segment durations $1/\tau_k$, so the oscillatory contributions from different segments do not add coherently. Second, because $JT\ll 1$, the relevant spin dynamics extends over many Floquet cycles ($n$ as the number of cycles). There is therefore a broad intermediate regime satisfying
$1/\omega \ll n\sum_{k\in\mathcal U_\alpha}\tau_k \ll 1/J,$ in which the fast motional modulation averages out while the spin dynamics remains slow. In this regime, thermal motion enters primarily through the shot-to-shot renormalization factors $\bar{\eta}_{ji}$, and Eq.~\eqref{eq:motion_factor} provides an effective description of the disorder in the couplings.

For the numerical simulations discussed below, we use the motional parameters: $\omega_r=2\pi\times59\,\mathrm{kHz}$, $\omega_z=2\pi\times13.3\,\mathrm{kHz}$, $\bar n_r=1.8$, and $\bar n_z=61$. These parameters determine the thermal distribution used to sample the disorder realizations entering the effective Hamiltonian.

\subsection{Single-Magnon Dynamics in the $1/r^3$ XXZ Model}
\subsubsection{Dispersion}
We first derive the single-magnon dispersion for the $1/r^3$ XXZ model in an infinite chain. In this case only the subspace with a single spin excitation, $\mathcal H_1^z$, is relevant. A convenient basis is $\ket{l}=\hat S_l^+\ket{0}$, where $\ket{0}=\ket{\downarrow\downarrow\cdots\downarrow}$ denotes the fully polarized state. A general state in this subspace can be written as
\begin{equation}
\ket{\psi}=\sum_l c_l\ket{l}.
\end{equation}
Acting with $\hat H_{\rm XXZ}$ gives
\begin{equation}
\sum_l\sum_r\left(\frac{J}{2r^3}c_{l-r}+\frac{J}{2r^3}c_{l+r}+E_0c_l-\varepsilon_1 c_l\right)\ket{l}=0,
\end{equation}
where $E_0$ is the constant energy offset from the Ising interaction. The eigenstates are plane waves,
\begin{equation}
\ket{\psi}=A\sum_l e^{ikl}\ket{l},
\end{equation}
which correspond to single-magnon excitations. Substituting this ansatz yields the dispersion relation
\begin{equation}
\varepsilon_1=E_0+J\sum_{r=1}^{\infty}\frac{\cos(rk)}{r^3}.
\end{equation}
The group velocity is therefore
\begin{equation}
v_g=\left|\frac{d\varepsilon_1}{dk}\right|=\left|J\sum_{r=1}^{\infty}\frac{\sin(rk)}{r^2}\right|.
\end{equation}
The Ising interaction enters only through the constant energy offset, so the dispersion and group velocity do not depend on $\Delta$. Numerically, the maximal group velocity is $v_g^{\rm max}=1.015J$.

\subsubsection{Quantum Walk Dynamics}\label{sec:quantum walk}
Experimentally, the simplest way to probe the single-magnon sector is to locally flip a single spin in the fully polarized state $\ket{0}$. This creates a localized excitation, which is a coherent superposition of single-magnon eigenstates with different quasi-momenta $k$. Under the XXZ Hamiltonian, this excitation undergoes a quantum walk along the chain.

To quantify the propagation speed, we compare the measured dynamics to a minimal nearest-neighbor tight-binding model. This model is not intended to reproduce all details of the long-range XXZ dynamics. Rather, it provides a simple and robust way to extract a single effective hopping parameter $t_{\rm eff}$ that characterizes the observed coherent quantum walk. We define the dimensionless quantity
\begin{equation}
\tilde t_{\rm eff}=\frac{2t_{\rm eff}}{J},
\end{equation}
which is used throughout the supplementary text.

\begin{figure}
\includegraphics{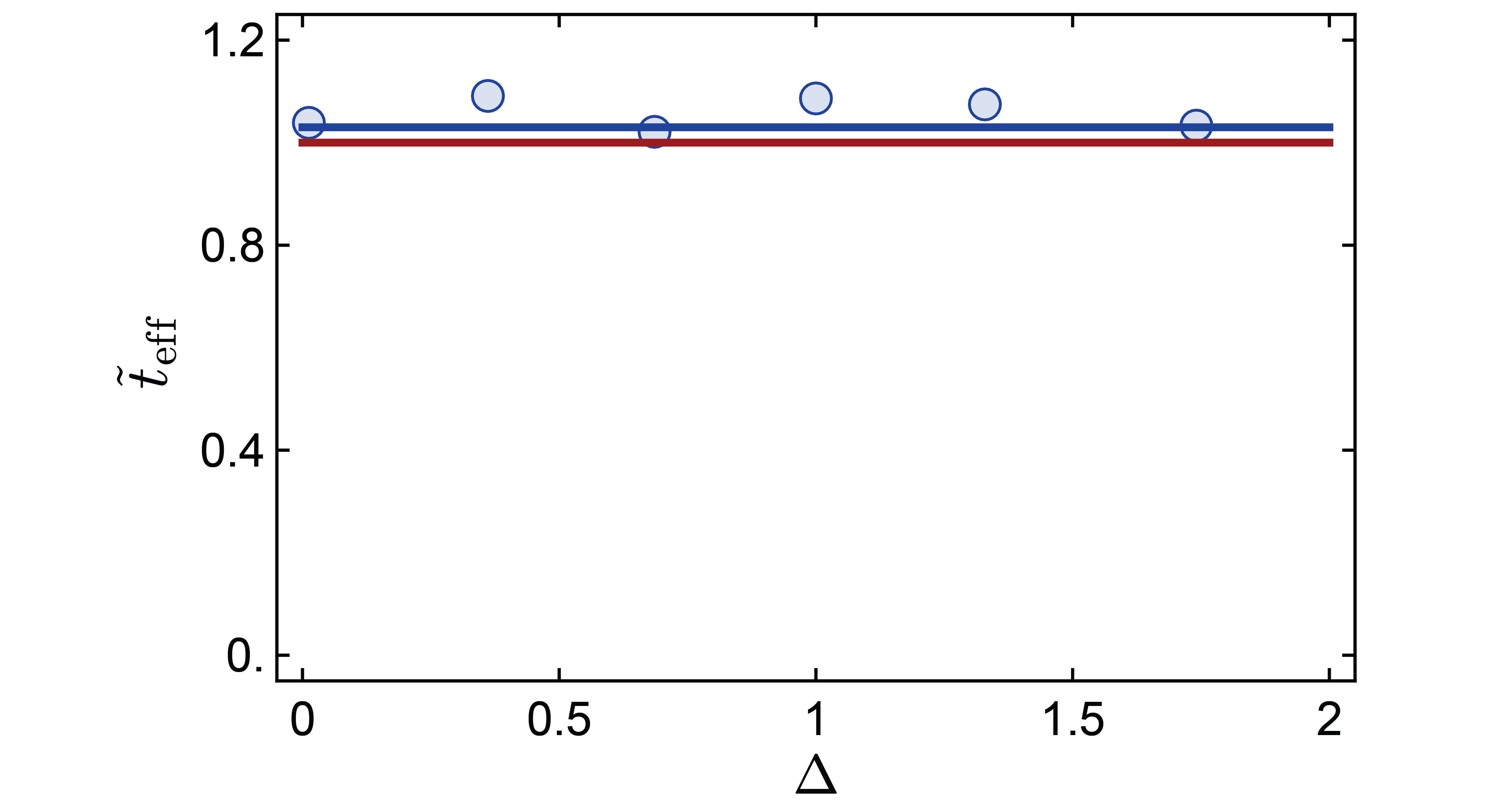}
\caption{ Normalized effective single-magnon hopping, $\tilde t_{\rm eff}=2t_{\rm eff}/J$, extracted from the quantum-walk fit as a function of $\Delta$. The blue curve shows the fitted value obtained from numerical simulations of a $1/r^3$ XXZ chain with $N=100$ sites, while the red curve shows the corresponding result for a nearest-neighbor XXZ chain. In both cases, $\tilde t_{\rm eff}$ is independent of $\Delta$.}
\label{fig:S6_single_magnon_speed}
\end{figure}

By mapping the spin operators to hard-core bosons,
\begin{equation}
\hat S_i^+\rightarrow b_i^\dagger, \qquad \hat S_i^-\rightarrow b_i,
\end{equation}
where $b_i^\dagger$ ($b_i$) creates (annihilates) a boson on site $i$ subject to the hard-core constraint $(b_i^\dagger)^2=b_i^2=0$, a flipped spin becomes a single localized hard-core boson. We then model the dynamics using the nearest-neighbor quantum-walk Hamiltonian
\begin{equation}
\hat H_{\rm QW}=t_{\rm eff}\sum_i\left(b_i^\dagger b_{i+1}+b_{i+1}^\dagger b_i\right),
\end{equation}
whose dispersion is
\begin{equation}
\varepsilon_{\rm QW}(k)=2t_{\rm eff}\cos k.
\end{equation}
The time-evolved state under $\hat H_{\rm QW}$ is
\begin{equation}
\ket{\psi_{\rm QW}(t)}=e^{-i\hat H_{\rm QW}t}\ket{\psi_0},
\end{equation}
where $\ket{\psi_0}$ is a single boson localized on the initially flipped site.

We can use this effective tight-binding model to extract propagation speeds of the local spin excitations, quantified by $\tilde{t}_{\text{eff}}$, from $\hat{S}^z_i$ measurements. For a seven-site chain, finite-size effects become important once the wave packet reaches the boundaries and reflects. In addition, the Ising term produces weak site-dependent diagonal energy shifts near the edges. To minimize the influence of these boundary effects, we restrict the fit to early-time dynamics, where the motion is dominated by bulk propagation. We extract $t_{\rm eff}$ by minimizing the least-squares cost function
\begin{equation}
\min_{t_{\rm eff}}\sum_{i,t}\left[\langle P_i^\uparrow(t)\rangle-\left|\langle i|\psi_{\rm QW}(t)\rangle\right|^2\right]^2.
\end{equation}
Uncertainties are estimated using bootstrap resampling: we generate multiple resampled datasets, fit each independently, and take the mean and standard deviation of the resulting $t_{\rm eff}$ values as the central value and error bar.

To benchmark the fitting procedure, we perform exact numerical simulations of a long XXZ chain with $N=100$ spins and compute $\langle P_i^\uparrow(t)\rangle$. Applying the same analysis yields $\tilde t_{\rm eff}\approx 1.03$ for the $1/r^3$ model, slightly faster than the nearest-neighbor XXZ chain which gives $\tilde t_{\rm eff}=1$. This confirms that $\tilde t_{\rm eff}$ provides a useful single-parameter characterization of the propagation speed in the long-range dipolar system. Applying the same procedure to the experimental data at several Ising anisotropies, we find that the extracted $\tilde t_{\rm eff}$ is approximately independent of $\Delta$ and consistent with the infinite-chain result (Fig.~\ref{fig:S6_single_magnon_speed}). Our data, which is systematically above $\tilde t_{\rm eff}=1$, hints at a small next-nearest-neighbor effect, consistent with theory.

\begin{figure}
\includegraphics{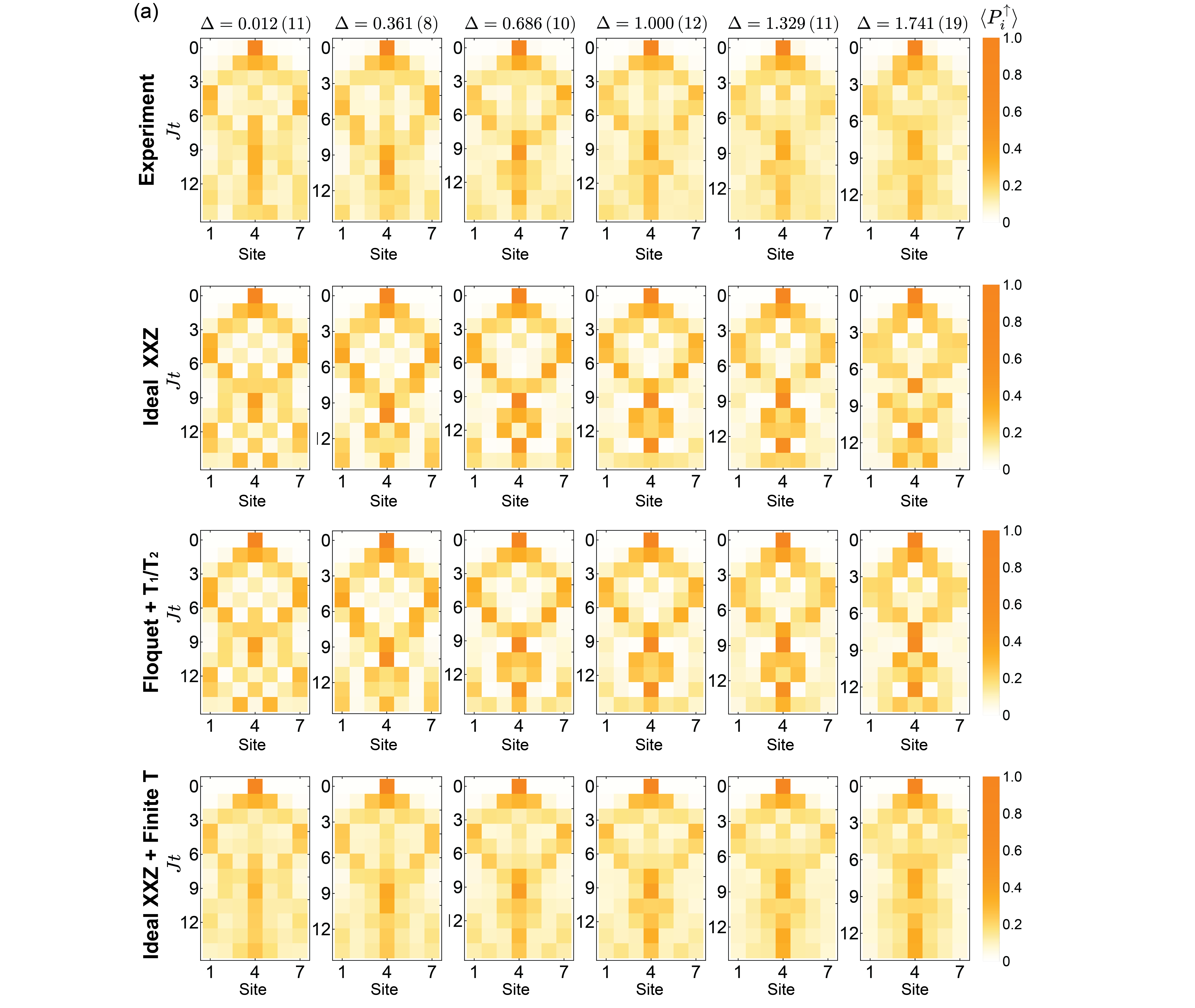}
\caption{\label{fig:S5_single_spin_QW} Quantum walk of single spin excitations. From top to bottom, we show $\langle \hat{S}^z_i\rangle$ for (i) experimental data, (ii) exact-diagonalization of the ideal $1/r^3$ XXZ model, (iii) simulations of the full Floquet engineering sequences with pulse errors, on-site disorder,  and $T_1$ and $T_2$ effects, (iv) simulations of the ideal $1/r^3$ XXZ models with finite-temperature induced interaction disorder.}
\vspace{-0.2in}
\end{figure}

\subsubsection{Reduction of Contrast in Quantum Walk}
We observe that the contrast of the single-spin quantum walk in Fig.~2(a) of the main text decreases for evolution times $t\gtrsim 10/J$. Our numerical analysis indicates that this damping is dominated by interaction disorder, rather than by higher-order Floquet terms or imperfect microwave control.

To assess the role of Floquet engineering and pulse imperfections, we first simulate the full Floquet engineering sequences in the complete Hilbert space. These simulations explicitly include the native spin-exchange Hamiltonian, the on-site disorder, and the microwave control pulses. The on-site disorder is modeled by a Gaussian distribution with width $\sigma_h=2\pi\times100\,\mathrm{Hz}$, chosen to match the single particle dephasing time $T_2^*$~\cite{Holland2023SE}. The microwave pulses are implemented explicitly in time, so that higher-order Floquet corrections are naturally included. To model pulse imperfections, we add Gaussian random amplitude and phase noise to each microwave pulse, with widths chosen to match the measured single-particle $T_1$ and $T_2$ values for the corresponding Floquet sequence. We find that, over the timescale relevant to Fig.~2(a) in the main text, these effects lead to negligible reduction of the quantum-walk contrast.

We then isolate the role of interaction disorder using the thermal-motion framework developed in Sec.~\ref{sec:Supp_ThermalMotion}. In this simulation, we do not explicitly include the Floquet pulse sequence. Instead, we evolve directly under the target $1/r^3$ XXZ Hamiltonian with shot-to-shot interaction disorder generated by thermally sampled molecular motion. This model reproduces the observed reduction of contrast remarkably well. Taken together, these results support the conclusion that interaction disorder is the dominant origin of the damping observed in the single-spin quantum walk (Fig.~\ref{fig:S5_single_spin_QW}).

\subsection{Two-Magnon Bound States in the $1/r^3$ XXZ Model}\label{sec:Supp_TwoMagnon}
\subsubsection{Dispersion and bound-state threshold}
\begin{figure}
	\includegraphics{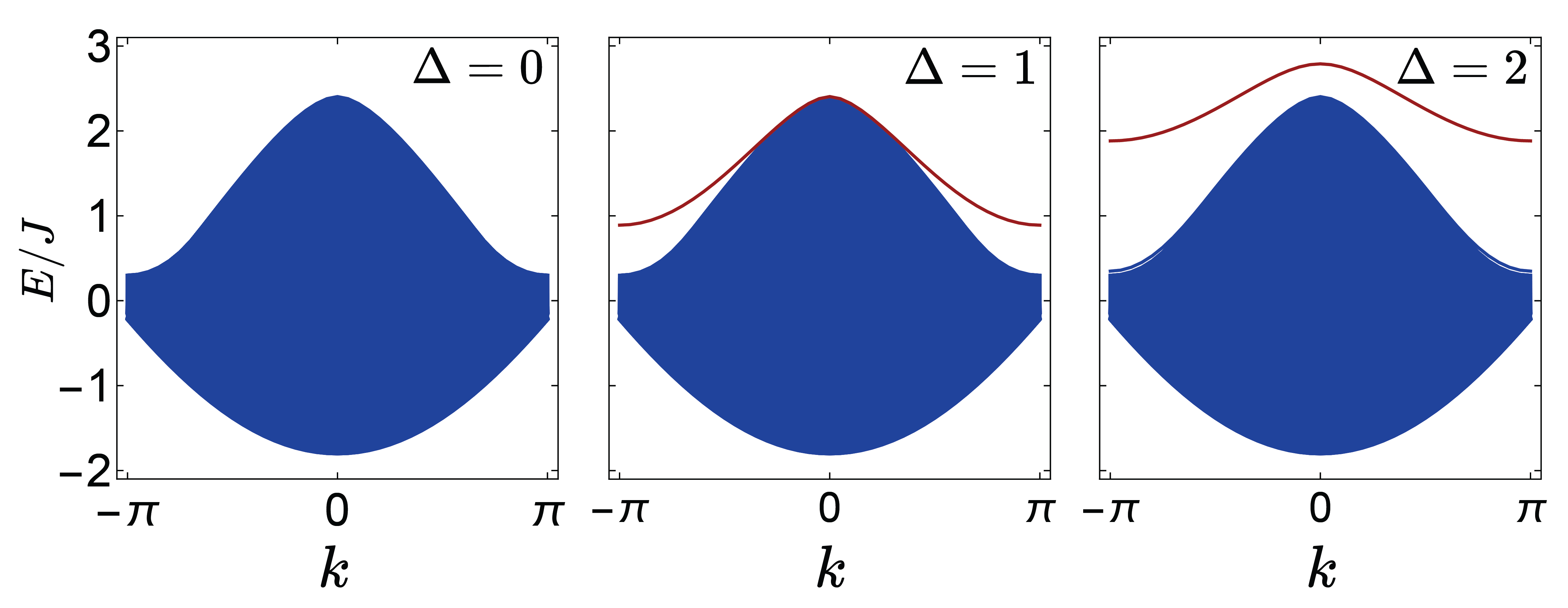}
	\caption{\label{fig:S7_BSDispersion} Two-magnon dispersion as a function of $\Delta$. The red lines indicate the isolated repulsive bound-state branch.}
\end{figure}

We characterize repulsive two-magnon bound states in the long-range XXZ model and relate them to the experimentally observed dynamics of two adjacent flipped spins. In the two-magnon sector $\mathcal H_2^z$, spanned by $\{\ket{i,j}\}_{i\neq j}$ with $\ket{i,j}=S_i^+S_j^+\ket{0}$, a general state has the form
\begin{equation}
\ket{\psi}=\sum_{i\neq j} c_{i,j}\,S_i^+S_j^+\ket{0}.
\end{equation}
Introducing relative and center-of-mass coordinates, $r=j-i\neq 0$ and $R=(i+j)/2$, and using the separation ansatz
\begin{equation}
c_{R,r}=e^{ikR}f(r), \qquad f(r)=f(-r), \qquad f(0)=0,
\end{equation}
one finds the relative-coordinate eigenvalue equation~\cite{Macri2021XXZBS}
\begin{align}
E\,f(r) &= \frac{J\Delta}{|r|^{\alpha}} f(r)
+J\sum_{\ell=1}^{\infty}\frac{\cos(k\ell/2)}{\ell^{\alpha}}\Big[f(r+\ell)+f(r-\ell)\Big],
\label{eq:Supp_rel_eq}
\end{align}
for $r\neq 0$, where the vacuum-energy offset from the Ising interaction has been subtracted. We truncate $r\in\{\pm1,\dots,\pm N_r\}$ and diagonalize the resulting $N_r\times N_r$ matrix with $N_r=200$, $\alpha=3$, and $J>0$. The resulting dispersion is shown in Fig.~\ref{fig:S7_BSDispersion}. As $\Delta$ increases, the highest-energy branch moves upward away from the two-magnon continuum and evolves into an isolated repulsive bound-state band.

To determine whether this branch remains isolated in the thermodynamic limit, we study the finite-size scaling of the separation between the upper branch and the non-interacting continuum. At fixed total momentum $k$, we define
\begin{equation}
\Delta E(N)=E_{\rm max}(k,\Delta;N)-E_{\rm max}(k,0;N),
\end{equation}
where $E_{\rm max}(k,\Delta;N)$ is the highest two-magnon eigenenergy for a system of length $N$. This quantity measures how far the repulsive branch lies above the free two-magnon continuum. We fit its finite-size dependence to
\begin{equation}
\Delta E(N)=\frac{a_0}{N^{a_1}}+\Delta E_\infty.
\label{eq:Supp_scaling_fit}
\end{equation}
A nonzero $\Delta E_\infty$ indicates that the upper branch remains separated from the continuum as $N\to\infty$, and therefore defines a fully isolated repulsive bound-state band.

For $\alpha=3$, we numerically find that the $k=0$ sector is the last to become isolated as $\Delta$ increases. We therefore use the $k=0$ gap as the criterion for the onset of a fully isolated band. Fitting the extracted $\Delta E_\infty$ to
\begin{equation}
A\,(\Delta-\Delta_c)^\nu\,\theta(\Delta-\Delta_c)
\end{equation}
yields $\Delta_c\simeq1.03$, where $\theta(x)$ is the Heaviside function.

\begin{figure}
	\includegraphics{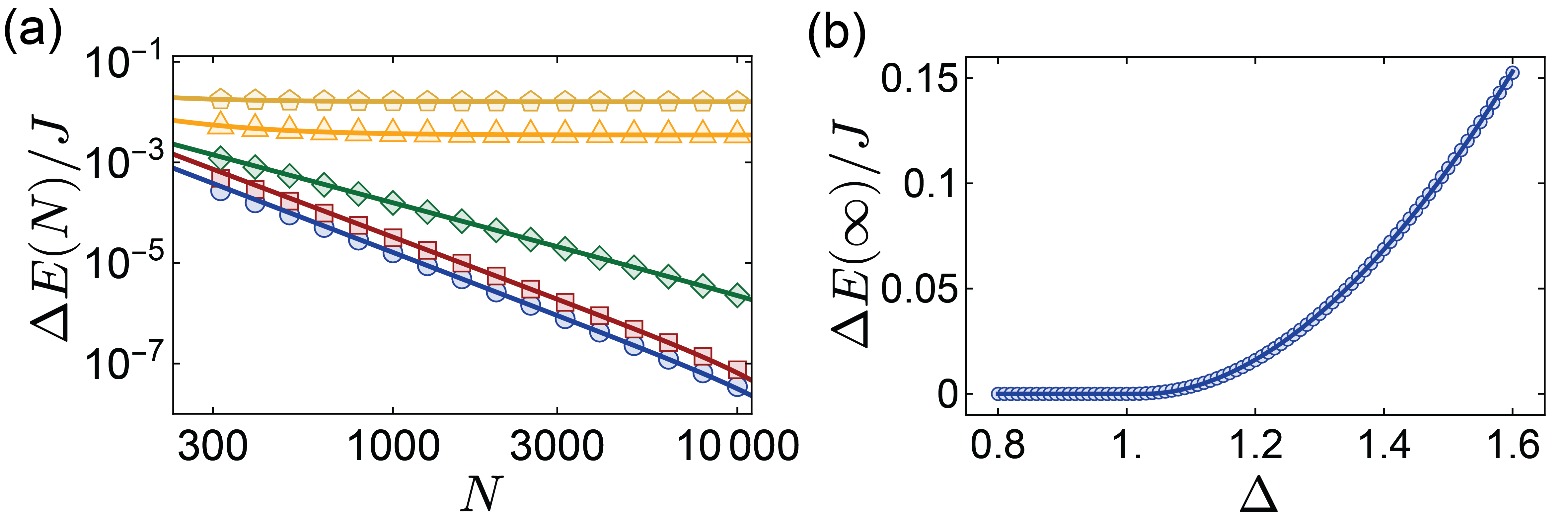}
	\caption{\label{fig:S8_BSThreshold} Finite-size scaling of the repulsive bound-state gap. (a) Gap $\Delta E(N)$ versus chain length $N$. Blue circles, red squares, green diamonds, orange triangles, and yellow pentagons correspond to $\Delta=0.8$, 0.9, 1.0, 1.1, and 1.2, respectively. Solid lines show fits to Eq.~\eqref{eq:Supp_scaling_fit}. (b) Thermodynamic-limit gap $\Delta E_\infty$ extracted from the fits in (a) as a function of $\Delta$.}
\end{figure}

\subsubsection{Experimental Observables for Probing the Magnon Bound State}
Experimentally, we prepare two adjacent flipped spins in an $N=8$ chain and monitor the subsequent dynamics of the system. This initial state is not a pure bound state; rather, it contains both continuum and bound-state two-magnon components. The continuum component tends to separate in real space, while the bound-state component remains concentrated at short relative distance.

To characterize the dynamics, we construct the distance distribution
\begin{equation}
P_b(d,t)=\sum_{i=1}^{N-d}\langle P_i^\uparrow P_{i+d}^\uparrow\rangle(t),
\label{eq:Supp_Pbd}
\end{equation}
which gives the probability that the two flipped spins are separated by distance $d$ at time $t$. In particular,
\begin{equation}
P_b(1,t)=\sum_{i=1}^{N-1}\langle P_i^\uparrow P_{i+1}^\uparrow\rangle(t)
\label{eq:Supp_Pb1}
\end{equation}
measures the total weight in the adjacent-pair subspace. The full two-point correlator $\langle P_i^\uparrow P_j^\uparrow\rangle(t)$ provides a site-resolved visualization of the two-magnon dynamics. For a bound state, correlations remain concentrated on the first off-diagonals, indicating that the two spin excitations propagate predominantly as an adjacent pair. As the Ising strength $\Delta$ increases, the bound state wavefunction becomes more and more localized to $d=1$ over all quasi-momenta.

We now explain why the adjacent-pair probability provides an experimentally useful probe of the bound-state fraction. Let $\Pi_{\rm BS}$ denote the projector onto the isolated repulsive bound-state band. We decompose the initial state as
\begin{equation}
\ket{\psi_0}=\ket{\psi_{\rm BS}}+\ket{\psi_{\rm cont}}, \qquad \ket{\psi_{\rm BS}}=\Pi_{\rm BS}\ket{\psi_0},
\end{equation}
with intrinsic bound-state weight
\begin{equation}
W_{\rm BS}=\mel{\psi_0}{\Pi_{\rm BS}}{\psi_0}.
\end{equation}
Defining the projector onto the adjacent-pair manifold,
\begin{equation}
\Pi_1=\sum_{i=1}^{N-1}\ket{D_i}\bra{D_i}, \qquad \ket{D_i}=\hat S_i^+\hat S_{i+1}^+\ket{0},
\end{equation}
one obtains
\begin{equation}
P_b(1,t)=\bra{\psi(t)}\Pi_1\ket{\psi(t)}=P_{\rm BS}(t)+P_{\rm cont}(t)+P_{\rm int}(t),
\end{equation}
where the three contributions arise from the bound-state component, the continuum component, and their interference. At short and intermediate times, two effects suppress the non-bound contribution to $P_b(1,t)$. First, the bound and continuum sectors acquire different dynamical phases, causing the interference term $P_{\rm int}(t)$ to dephase. Second, the continuum component rapidly loses support in the adjacent-pair sector as the two excitations separate in real space, causing $P_{\rm cont}(t)$ to decay. Before boundary reflections become important, this motivates the empirical fit form
\begin{equation}
P_b(1,t)\approx A e^{-\kappa t}+P_B,
\label{eq:PB_fit}
\end{equation}
where $P_B$ is the equilibrium value once the continuum contribution has largely left the adjacent-pair sector.

More explicitly, after the initial dephasing transient one expects
\begin{equation}
P_B\approx \eta_{\rm BS}W_{\rm BS},
\end{equation}
where
\begin{equation}
\eta_{\rm BS}=\frac{\sum_{\mu\in{\rm BS}} |a_\mu|^2 \mel{\mu}{\Pi_1}{\mu}}{\sum_{\mu\in{\rm BS}} |a_\mu|^2},
\qquad
\ket{\psi_{\rm BS}}=\sum_{\mu\in{\rm BS}} a_\mu \ket{\mu}.
\end{equation}
For a tightly bound branch, $\eta_{\rm BS}\to 1$, and $P_B$ directly tracks the bound-state fraction. In practice, we extract $P_B$ by fitting Eq.~\eqref{eq:PB_fit} over the time window $t\lesssim 6/J$, before boundary reflections become significant. The observed increase of $P_B$ with $\Delta$ is therefore consistent with the increasing overlap of the initial state with the isolated repulsive bound-state band.

Lastly, the coherent motion within the adjacent-pair sector can be accessed through the post-selected center-of-mass distribution
\begin{equation}
P_{{\rm CM},i}(t)=\frac{\langle P_i^\uparrow P_{i+1}^\uparrow\rangle(t)}{\sum_j \langle P_j^\uparrow P_{j+1}^\uparrow\rangle(t)},
\label{eq:PCM}
\end{equation}
which is the probability that the two flipped spins occupy bond $i$, conditioned on their remaining adjacent.

\subsubsection{Monte Carlo Modeling of the Bound State Proxy $P_B$}\label{sec:Supp_ThermalBSFrac}
We find that the measured bound state proxy $P_B$ deviates from simulation results of the ideal $1/r^3$ XXZ model. We attribute this primarily to interaction disorder, with additional contributions from imperfect state preparation and detection. To capture those effects, we perform the following Monte Carlo simulations.

For each simulated shot, we first sample the initial site occupancy and spin configuration based on characterized imperfect state preparation fidelities. We then sample the motional coordinates from the measured thermal distribution and construct the corresponding shot-dependent couplings $\bar{\eta}_{ij}$, which define a disordered effective XXZ Hamiltonian. In addition, we include the experimental tweezer compression and decompression ramps, during which no Floquet drive is applied and the system evolves under the native XY interaction with Gaussian on-site disorder. After simulating the spin evolution, we apply readout errors and process the resulting snapshots using the same post-selection and fitting procedures as in the experiment.

This model also explains why the initial value, $P_b(d=1,t=0)$, is less than unity in the experiment. Even at zero nominal interaction time, the initial adjacent pair acquires some spreading during the array ramps, because the molecules evolve under the bare exchange Hamiltonian while the tweezers are moved into and out of the interacting geometry. From the simulated snapshots, we compute $P_b(d,t)$ and extract $P_B$ using the same exponential fitting function as used in analyzing the experimental data. The resulting curves are shown in Fig.~3(c) of the main text. We find that thermal motion together with state-preparation and readout imperfections largely accounts for the observed suppression of $P_B$ relative to the ideal exact-diagonalization result, while preserving the qualitative feature that it increases with Ising strength $\Delta$, signaling the emergence of magnon bound states.

\subsubsection{Coherent Quantum Walk of the Two-Magnon Bound State}\label{sec:Supp_TwoMagnon_QW}
The post-selected distribution $P_{{\rm CM},i}(t)$ in Eq.~\eqref{eq:PCM} isolates the coherent dynamics within the adjacent-pair sector. For the values $\Delta=1.329(11)$ and $1.741(19)$ explored experimentally, the repulsive bound-state band is already well isolated, and $P_{{\rm CM},i}(t)$ exhibits a clear coherent quantum walk of the bound pair.

This behavior can be understood perturbatively in the strong-Ising regime ($\Delta\gg 1$). We partition the Hamiltonian as
\begin{equation}
\hat H_{\rm XXZ}=\hat H_0+\hat H',
\end{equation}
with
\begin{equation}
\hat H_0 = \sum_{i<j}\frac{J\Delta}{|i-j|^3}\hat S_i^z\hat S_j^z, \qquad \hat H' = \sum_{i<j}\frac{J}{|i-j|^3}\,\frac{1}{2}(\hat S_i^+\hat S_j^-+\hat S_i^-\hat S_j^+).
\end{equation}
To second order in $\hat H'$, the effective Hamiltonian within the adjacent-pair manifold is
\begin{equation}
\hat H_{\rm eff}^D = \mathcal P\hat H'\mathcal P + \mathcal P\hat H'\mathcal Q\frac{1}{E_1^{(0)}-\mathcal Q\hat H_0\mathcal Q}\mathcal Q\hat H'\mathcal P + \mathcal O(\hat H'^3),
\label{eq:Supp_SW}
\end{equation}
with projectors
\begin{equation}
\mathcal P=\sum_i \ket{D_i}\bra{D_i}, \qquad \mathcal Q=\mathbb I-\mathcal P.
\end{equation}
Here $E_1^{(0)}$ is the unperturbed energy of an adjacent pair. After subtracting the vacuum-energy offset, the unperturbed energy of a two-magnon product state depends only on the separation $d$,
\begin{equation}
E_d^{(0)}=\frac{J\Delta}{d^\alpha}.
\end{equation}

The effective hopping of the bound pair has two distinct contributions. The first is a direct first-order process arising from long-range exchange:
\begin{equation}
\mel{D_{i+1}}{\hat H'}{D_i}=\frac{J}{2}\frac{1}{2^\alpha}.
\end{equation}
The second is a second-order process in which the pair is virtually broken and then reformed one bond away. Summing over intermediate states in $\mathcal Q$ gives
\begin{equation}
\begin{aligned}
t_{\rm eff}^{(2)}=\frac{J}{4\Delta}\bigg[\frac{1}{1-2^{-\alpha}} + 2\sum_{d=2}^{\infty}\frac{1}{(d+1)^\alpha(d-1)^\alpha(1-d^{-\alpha})}\bigg].
\end{aligned}
\end{equation}
Keeping only nearest-neighbor hopping in the adjacent-pair manifold, we obtain
\begin{equation}
\hat H_{\rm eff}^D\approx \sum_i t_{\rm eff}\left(\hat D_i^+\hat D_{i+1}^-+\hat D_i^-\hat D_{i+1}^+\right),
\end{equation}
with
\begin{equation}
t_{\rm eff}=\frac{J}{2^{\alpha+1}}+t_{\rm eff}^{(2)}.
\label{eq:teff}
\end{equation}
This makes clear why the $1/r^3$ model differs qualitatively from a nearest-neighbor XXZ chain: in addition to the usual second-order pair motion proportional to $1/\Delta$, the long-range model has a direct first-order hopping channel.

In the data analysis, we fit the measured $P_{{\rm CM},i}(t)$ to a nearest-neighbor tight-binding quantum walk model on the bond lattice and extract the corresponding $t_{\rm eff}$, following the same procedure used for the single-magnon dynamics. To benchmark this procedure, we first extract $t_{\rm eff}$ from exact-diagonalization results for XXZ chains with $N=20$ sites and find that, in the strong-Ising regime, the fitted values agree well with Eq.~\eqref{eq:teff}. For the experimental system size $N=8$, finite-size effects are non-negligible, so we compare the extracted values directly with exact diagonalization for the finite chain. As shown in Fig.~3(g) of the main text, the measured values at $\Delta=1.329(11)$ and $1.741(19)$ agree with the $1/r^3$ XXZ model and deviate strongly from a nearest-neighbor model, supporting our interpretation of the post-selected dynamics as the coherent propagation of a repulsively bound magnon pair.

\subsection{Reduction of Contrast in the Experimental $1/r^3$ XYZ Dynamics}
\subsubsection{Depolarization and parity dynamics}
A key feature of the XYZ Hamiltonian is the pair-creation term
\begin{equation}
\hat H_{\rm pair}=\sum_{i<j}\frac{\gamma J}{2|i-j|^3}(\hat S_i^+\hat S_j^++\hat S_i^-\hat S_j^-),
\end{equation}
which breaks conservation of $\hat S^z$ while preserving the parity of the number of spin-up excitations. Starting from the fully polarized state $\ket{\downarrow}^{\otimes N}$, the ideal dynamics therefore remain entirely within the even-parity sector. In the experiment, however, imperfections lead to leakage into the odd-parity sector, as seen in the full counting statistics.

We first observe that the parity operator is a global operator involving all sites. Thus, one expects that the decay rate due to experimental imperfections will generically be $N_{\text{mol}}$ faster than single-particle/local observables. We next explicitly show this in a rate equation model. To quantify deviations from perfect parity conservation, we define the total even- and odd-parity populations,
\begin{equation}
P_e(t)=\sum_{N_\uparrow\,\mathrm{even}} P(N_\uparrow,t), \qquad P_o(t)=\sum_{N_\uparrow\,\mathrm{odd}} P(N_\uparrow,t),
\end{equation}
and the parity contrast
\begin{equation}
C_P(t)=P_e(t)-P_o(t).
\end{equation}
Under ideal XYZ dynamics, $C_P(t)=1$ at all times, whereas in the experiment it decays approximately exponentially.

We first consider a single molecule subject to incoherent bit-flip errors at rate $\Gamma_1$. The populations obey
\begin{equation}
\frac{dP_\uparrow}{dt}=-\Gamma_1P_\uparrow+\Gamma_1P_\downarrow,
\end{equation}
\begin{equation}
\frac{dP_\downarrow}{dt}=-\Gamma_1P_\downarrow+\Gamma_1P_\uparrow,
\end{equation}
so that
\begin{equation}
P_d=P_\uparrow-P_\downarrow=e^{-2\Gamma_1 t}.
\end{equation}
The corresponding single-particle depolarization time is therefore $T_1=1/(2\Gamma_1)$, as described above.

We now extend this discussion to $N_{\rm mol}$ molecules. For a given molecule, let $q(t)$ be the probability that it has undergone an odd number of bit flips by time $t$. Since each additional flip toggles this parity, $q(t)$ obeys
\begin{equation}
\frac{dq}{dt}=\Gamma_1(1-q)-\Gamma_1 q=\Gamma_1(1-2q),
\end{equation}
with $q(0)=0$, giving
\begin{equation}
q(t)=\frac{1-e^{-2\Gamma_1 t}}{2}.
\end{equation}
Accordingly, the average parity contribution of a single molecule is
\begin{equation}
1-2q(t)=e^{-2\Gamma_1 t}.
\end{equation}
Assuming that bit flips on different molecules are independent, the total parity contrast is the product of the single-molecule parity factors,
\begin{equation}
C_P(t)=\prod_{i=1}^{N_{\rm mol}} [1-2q_i(t)] = e^{-2N_{\rm mol}\Gamma_1 t}.
\end{equation}
Thus the characteristic parity-decay time is
\begin{equation}
\tau = \frac{1}{2N_{\rm mol}\Gamma_1}=\frac{T_1}{N_{\rm mol}}.
\end{equation}

For this comparison, the relevant quantity is the $T_1$ value without blackbody correction. For the XYZ model with $\Delta=1.033(11)$ and $\gamma=0.466(5)$, we measure $T_1=150(12)/J_{\rm ex}$, while the experimentally observed parity-decay time is $\tau=9.6(8)/J_{\rm ex}$, faster than the estimate based on simple independent bit flips. This indicates additional parity-changing mechanisms beyond uncorrelated single-particle depolarization. Possible sources include temporally correlated noise from local on-site disorder due to differential light shifts, which can break the exact time-reflection symmetry of the Floquet cycle and generate higher-order parity-non-conserving terms, as well as spatially correlated errors that induce multi-spin flips. More generally, because parity is a global many-body observable, it is sensitive to many distinct parity-violating error channels, making the microscopic origin of the observed decay difficult to identify.

\subsubsection{Dynamics of Pair Creation/Annihilation and Spin-Exchange Correlators}
Here, we discuss additional spatial correlation data of pair creation/annihilation and spin-exchange. We prepare spin-polarized states $\ket{\downarrow}^{\otimes N_{\rm mol}}$ and measure two connected types of correlators, the hopping correlator
\begin{equation}
C^h_{ij}= (\langle S_i^+S_j^- +  S_i^-S_j^+\rangle_c )/2= \langle S_i^xS_j^x\rangle_c +  \langle S_i^yS_j^y\rangle_c,
\end{equation}
and the pairing correlator
\begin{equation}
C^p_{ij}=(\langle S_i^+S_j^+ +  S_i^-S_j^-\rangle_c)/2 = \langle S_i^xS_j^x\rangle_c -  \langle S_i^yS_j^y\rangle_c,
\end{equation}
where $\langle AB\rangle_c=\langle AB\rangle-\langle A\rangle\langle B\rangle$ denotes the connected correlator. The correlator $C^h_{ij}$ probes the spin-exchange processes and encodes the spin-exchange energy contribution to the system;  $C^p_{ij}$ probes correlations of pair creation and annihilation, and encodes the energy contribution of these processes. 
Notably, these correlators are easily accessible in our platform through measuring spatial correlations in the $x$ and $y$ bases. These, along with the correlators of $\hat{S}^z_i \hat{S}^z_j$ corresponding to the Ising energy term that are also accessible, allow full characterization of the energy interplay between all three terms in the XYZ Hamiltonian.  

\begin{figure}
\includegraphics[width=\textwidth]{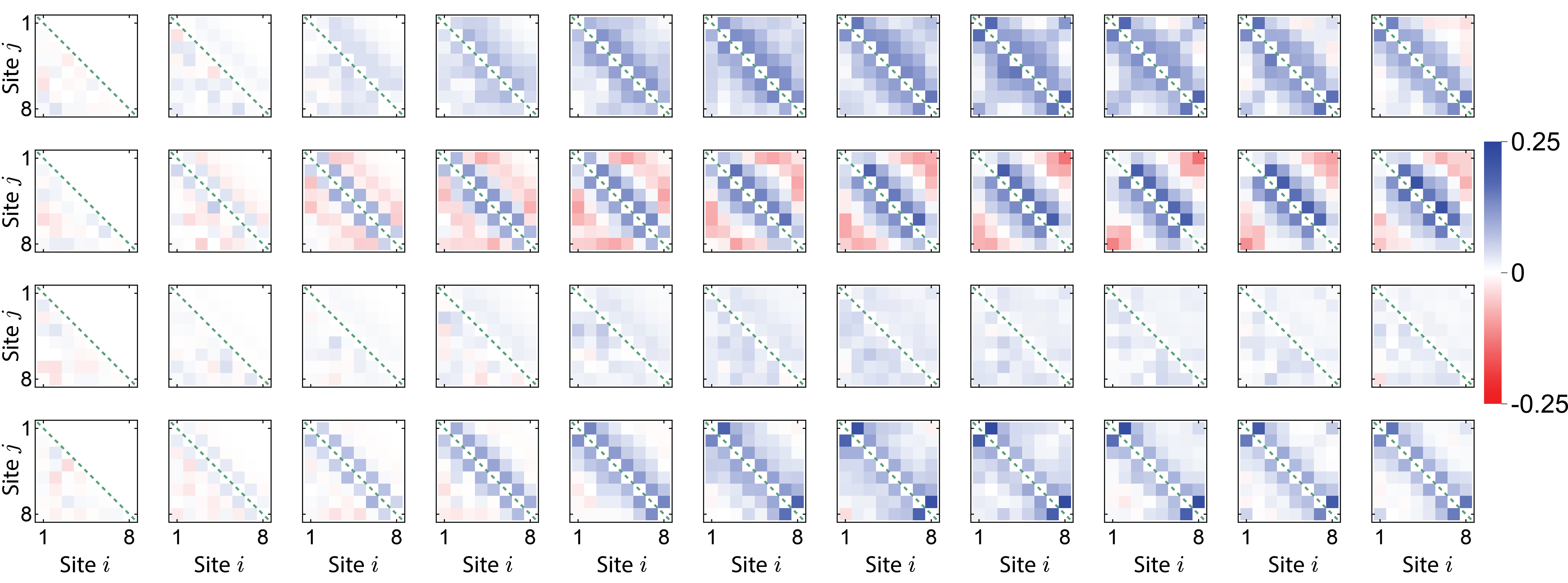}
\caption{\label{fig:S9_xyz_correlator} Dynamics of the hopping correlator $C_{ij}^{h}$ and pairing correlator $C_{ij}^{p}$ for two Floquet-engineered XYZ Hamiltonians. The top two rows show the dynamics of $C_{ij}^h$ and $C_{ij}^p$ for $H_{\rm XYZ,1}$ with $J=2\pi\times22.17(12)$ Hz, $\Delta=1.033(11)$, and $\gamma=0.466(5)$, while the bottom two rows show the corresponding dynamics for $H_{\rm XYZ,2}$ with $J=2\pi\times17.22(10)$ Hz, $\Delta=1.860(13)$, and $\gamma=0.290(3)$. From left to right, the evolution time increases from 0 to 39 ms in steps of 3.9 ms. In each panel, the upper triangle shows exact-diagonalization results for the ideal XYZ model, while the lower triangle shows the corresponding experimental data. Data in this figure are not post-selected for even parity, which is not possible since the correlators come from measurements in the $\hat{S}^x_i$ and $\hat{S}^y_i$ basis.}
\vspace{-0.2in}
\end{figure}

We first study the XYZ model with Ising strength $\Delta=1.033(11)$ and pair-creation strength $\gamma=0.466(5)$. As shown in Fig.~\ref{fig:S9_xyz_correlator}, the hopping correlator $C^h_{ij}$ initially develops at short distances and then spreads to longer ranges with time. Its magnitude reaches values up to $\sim 0.17$ and remains positive. At the same time, the pair correlator $C^p_{ij}$ exhibits similar dynamics and a comparable overall magnitude, but takes both positive and negative values. In particular, significant negative correlations develop between the two edges of the chain, indicating that it is energetically favorable to flip pairs of spins on opposite ends.

We next study the XYZ model with stronger Ising interaction $\Delta=1.860(13)$ and weaker pair creation $\gamma=0.290(3)$. In this regime, the hopping correlator $C^h_{ij}$ is strongly suppressed, reaching a maximum mean strength of only $\sim 0.03$. This can be understood as a consequence of stronger magnon binding at larger $\Delta$. Within a nearest-neighbor picture, starting from a spin-polarized state, pair creation initially generates neighboring $\ket{\uparrow\uparrow}$ excitations. If these excitations propagate primarily as a magnon bound pair, their motion is governed by the slower effective hopping of the composite object, which decreases with increasing $\Delta$. By contrast, the pairing correlator $C^p_{ij}$, although somewhat reduced, remains sizable. Overall, the measured correlators agree well with exact-diagonalization simulations. They demonstrate that our platform can probe not only the distinct dynamical effects of the different terms in $\hat H_{\rm XYZ}$, but also the nontrivial spatial correlations they generate.

